\documentstyle[12pt]{article}
\newcommand{\eq}{\begin{equation}}
\newcommand{\en}{\end{equation}}
\newcommand{\BEF}{\begin{figure}}
\newcommand{\EF}{\end{figure}}
\newcommand{\bea}{\begin{eqnarray}}
\newcommand{\eea}{\end{eqnarray}}

\newcommand{\th}{\theta}

\newcommand{\partiall}{\partial\hspace{-6pt}/}
\newcommand{\kbar}{k\hspace{-6pt}/}
\newcommand{\A}{A\hspace{-6pt}/}

\newcommand{\inta}{\int d^{2}x\sqrt{g}\,}

\newcommand{\lb}{\lambda}
\newcommand{\J}{\J_{\mu ab}}
\newcommand{\T}{\T^{\mu\nu}}

\newcommand{\semi}{;\hfil\break}

\newcommand{\N}{A^{(n)}}

\newcommand{\unity}{1\kern-.65mm \mbox{\form l}}
\newfont{\form}{cmss10}

\begin{document}
\setlength{\baselineskip}{12pt}
\begin{flushright}
November 1994

DFPD 94/TH/62
\end{flushright}

\vspace{2cm}
\begin{center}
{\Large\bf
The generalized chiral Schwinger model on the two-sphere}

\vspace{1cm}
{\bf A. Bassetto}\\
{\it Dipartimento di Fisica ``G.Galilei", Via Marzolo 8 --35131 Padova,
Italy\\
and INFN, Sezione di Padova, Italy\\}
\vspace{0.5cm}
{\bf L. Griguolo}\\
{\it International School for Advanced Studies\\
via Beirut 2, 34100 Trieste, Italy\\
INFN Sezione di Trieste\\}
\end{center}

\vspace{1cm}
\begin{abstract}

A family of theories
which interpolate between vector and chiral
Schwinger models is studied on the two--sphere $S^{2}$.
The conflict between the loss of gauge
invariance and global geometrical properties is solved
by introducing a fixed background connection.
In this way the generalized Dirac--Weyl operator can be globally defined
on $S^{2}$. The generating functional
of the Green functions is obtained by taking carefully
into account the contribution
of gauge fields with non--trivial topological charge and of the
related zero--modes of the Dirac determinant. In the decompactification
limit, the Green functions of the flat case are recovered; in
particular the fermionic condensate in the vacuum vanishes,
at variance with its behaviour in the
vector Schwinger model.

\end{abstract}

\newpage

\section{Introduction}

Quantum field theories in 1--space, 1--time dimensions are intensively
studied in recent years owing to their peculiarity of being exactly
solvable both by functional and by operatorial techniques.
\noindent
{}From a practical point of view they find interesting applications in string
models, while behaving as useful theoretical laboratories in which many
features, present also in higher dimensional theories, can be directly
tested. In addition 2--dimensional models possess a quite peculiar infrared
structure on their own.

Historically the first 2--dimensional model was proposed by
Thirring \cite{Thi58}, describing a pure fermionic current--current
interaction.
The interest suddenly increased 4 years later, when Schwinger
\cite{Sch62} was able
to obtain an exact solution for 2--dimensional electrodynamics with
massless spinors.

Chiral generalizations of this model were studied by Hagen
\cite{Hag73} and, more recently, by Jackiw and
Rajaraman \cite{Jac85}. The last authors draw very important
conclusions concerning
theories with ``anomalies", i. e. the occurrence of symmetry breakings by
quantum effects \cite{uno}, \cite{due}, \cite{tre}, \cite{quattro}.
They were able to show that, taking advantage of the arbitrariness in the
(non perturbative) regularization of the fermionic determinant, it was
possible to recover a unitary theory even in the presence of a gauge
anomaly (gauge non--invariant formulation of an anomalous gauge theory).

In ref.\cite{Gri94} a family of theories
which interpolate between vector and chiral
Schwinger models according to a parameter {\it r}, which tunes the ratio of the
axial to vector coupling, has been studied.
We call it generalized chiral Schwinger model.
\noindent
The treatment depends on two parameters: {\it r} and {\it a},
{\it a} being the
constant involved in the regularization of the fermionic determinant.

In the Minkowski space, using a non--perturbative
approach, we first obtained, by means of a functional formalism, the
correlation functions for bosons, fermions and fermionic composite
operators.
We found two allowed ranges for the parameters {\it r} and {\it a}.
The first range was also partially studied in a similar context in
\cite{Hal86}, \cite{Miy88}. In this range
the bosonic sector consists of two ``physical" quanta, a free massive
and a free massless excitation. The fermionic sector is much more
interesting: both left and right spinors exhibit a propagator decreasing at
very large distances, indicating the presence of asymptotic states which
however feel the long range interaction mediated by the massless boson.
The asymptotic fermions are described by a massless Thirring model.

The solution interpolates between two conformal invariant theories at small
and large distances, respectively, with different critical exponents.
The c--theorem \cite{Zam86} is explicitly verified, confirming
that $\Delta c=1$, as one could expect on the basis of the structure of the
bosonized theory.

The second range is characterized in the bosonic sector by a ``physical"
massive excitation and by a massless negative norm state (``ghost").
Both quanta are free; one can define a stable Hilbert space of states in
which the ``ghost" does not appear. However no asymptotic states for
fermions are available in this case; their correlation function increases
with distance, giving rise to a confinement phenomenon.

All those features were confirmed and further elucidated by a treatment
based on
operators which are canonically quantized according to a Dirac bracket
formalism \cite{Dir50}.

A perturbative approach to
this generalized chiral Schwinger model
has also been proposed in ref.\cite{Gro94}.

First the perturbative
expansion for the boson propagator is resummed, starting from the Feynman
diagrams: in order to develop the Feynman rules we had to introduce a
gauge fixing.

In the non--perturbative context, where gauge invariance is naturally broken
by the anomaly, this amounts to studying different theories for different
gauge fixings. Therefore the limit of vanishing gauge fixing was performed
after resummation. A lot of interesting features was hidden in
this limit:
studying the bosonic spectrum, the decoupling of ghost
particles from the theory was followed, to
recover the previous non perturbative results.

The fermionic correlation functions were also examined, leading to
the correct Thirring behaviour in the non--perturbative limit; nevertheless
we found very different ultraviolet scaling laws before and after the
gauge--fixing removal, related to the appearance of an ultraviolet
renormalization constant.
Decoupling of heavy states is indeed not trivial when anomalies
are present \cite{Fa84}.

In this paper we study the euclidean version of this generalized chiral
Schwinger model on the two--sphere $S^{2}$:
the problem is of interest
by itself because there is a conflict between the loss of gauge
invariance  and the globality properties of the model.

It is well known
that gauge anomalies in presence of non--trivial fiber--bundle depend on
some ``fixed'' background connection \cite{Stora85}: the global meaning
of the cohomological solution requires the presence of these
connections. From the functional point of view we will show that the
determinant of the generalized Dirac--Weyl operator is globally defined
on $S^{2}$ only after the introduction of a classical external field.

We discuss its physical meaning and we obtain the generating functional
of the Green functions: the contribution
of gauge fields with non--trivial topological charge and of the
related zero--modes of the Dirac determinant is carefully
taken into account. As an
application we derive the fermionic
propagator and the condensate in the vacuum: we find that, in the
decompactification limit, the latter
vanishes, at variance with the
Schwinger model case, confirming the conjecture \cite{Gir86} about the
triviality of the vacuum of the chiral Schwinger model.

In sect.2 we recall the basic results of \cite{Gri94}, following
the path-integral approach, and establish our notations.

Sect.3 deals with the geometrical background necessary to properly
define and treat the Dirac-Weyl determinant, which is studied
in sect.4.
The chiral gauge theory on the sphere is defined in sect.5
and its generating functional is constructed in sect.6.
The
fermionic two-point function and the vacuum expectation value for the
scalar density are computed and discussed in sect.7, and the
decompactification limit is performed.
Conclusions are drawn in sect.8.

\vfill\eject

\section{The non--perturbative solution of the generalized
chiral Schwinger model}

\par
The model, characterized by the classical lagrangian
\begin{equation}
\label{uno}
{\cal L} = -{1 \over 4} F_{\mu \nu} F^{\mu \nu}+ \bar \psi\,\gamma^{\mu}
\Bigl[\,i
\partial_{\mu} +
e \,({1 +r \gamma_{5} \over 2}) A_{\mu}\,\Bigr] \psi,
\end{equation}
\noindent
is quantized following the path--integral method. In
eq.(\ref{uno}) $F_{\mu\nu}$ is the usual field tensor, $A_\mu$ the vector
potential and $\psi$ a massless spinor. The quantity $r$ is a real
parameter interpolating between the vector $(r=0)$ and the chiral $(r=\pm
1)$ Schwinger models. Our notations are
\begin{eqnarray}
\label{due}
g_{00}=- g_{11}&=& 1, \qquad \epsilon^{01}= -\epsilon_{01} =1, \nonumber\\
\gamma^0&=&\sigma_1 ,\qquad \gamma^1=- i\sigma_2, \nonumber\\
\gamma_5&=&\sigma_3, \qquad
\tilde \partial_\mu = \epsilon_{\mu \nu} \partial^\nu,
\end{eqnarray}
\noindent
$\sigma_i$ being the usual Pauli matrices.

The classical lagrangian
eq.(\ref{uno}) is invariant under the local transformations

\bea
\psi'(x)&=&\exp\Bigl[ie{(1+r\gamma^5)\over 2}\Lambda(x)\Bigr]\psi(x)
\nonumber\\
A'_\mu(x)&=&A_\mu(x)+\partial_\mu\Lambda.
\label{23}
\eea
However, as is well known, it is impossible to make the fermionic
functional measure simultaneously invariant under  vector and axial vector
gauge transformations; as a consequence, for $r\neq 0$ the quantum theory
will exhibit anomalies.

The Green function generating functional is
\eq
{\cal Z}[J_\mu,\bar\eta, \eta]={\cal Z}_0^{-1}\int{\cal D}(A_\mu, \bar\psi,
\psi)\exp\Bigl[\,i\int\,d^2x({\cal L}+{\cal L}_s)\,\Bigr],
\label{24}
\en
\noindent
and
\eq
{\cal L}_{s}=J_\mu A^\mu+\bar\eta\psi+\bar\psi\eta
\label{25}
\en
\noindent
$J_\mu$, $\eta$ and $\bar\eta$ being vector and spinor sources
respectively.

The integration over the fermionic degrees of freedom can be
performed, leading to the expression
\bea
{\cal Z}[J_\mu, \eta, \bar\eta]&=&{\cal Z}_0^{-1}
\int{\cal D}(A_\mu,\phi)\exp\Bigl[i\int d^2x
{\cal L}_{eff}(A_\mu, \phi)+d^2 xJ_\mu A^\mu\Bigr]\nonumber\\
& &\exp\Bigr[-i\int
d^2x\,d^2y\,\bar\eta(x)S(x,y;A_\mu)\,\eta(y)\Bigr]
\label{26}
\eea
\noindent
where
\eq
{\cal L}_{eff}=-{1\over 4}F_{\mu\nu}F^{\mu\nu}+{ae^2\over 8\pi}A_\mu
A^\mu+{1\over 2}\partial_\mu\phi\partial^\mu\phi+{e\over
2\sqrt{\pi}}A^\mu(\tilde\partial_\mu-r\partial_\mu)\phi,
\label{27}
\en
\noindent
$\phi$ being a scalar field we have introduced in order to have a local
${\cal L}_{eff}$ and {\it a} the subtraction parameter reflecting the
well--known regularization ambiguity of the fermionic determinant
\cite{Jac85}.

The quantity $ S(x,y;A_\mu)$ in eq.(\ref{26}) is the fermionic propagator
in the presence of the potential $A_\mu$, which will be computed later on
by using standard decoupling techniques.

For the moment we let the sources $\eta$ and $\bar\eta$ vanish and consider
the bosonic sector of the model for different values of the parameters $r$
and $a$. In this sector the effective lagrangian is quadratic in the
fields; this means an essentially free (although non local) theory.
First functionally integrating over $\phi$ and then over $A_\mu$, we easily
obtain
\eq
{\cal Z}[J_\mu, 0, 0]=\exp [-{1\over 2}\int d^2
xJ^\mu(K^{-1})_{\mu\nu}J^\nu ],
\label{28}
\en
\noindent
where
\eq
K_{\mu\nu}=g_{\mu\nu} (\Box + {e^2 \over 4\pi}(1+a)-(1+
{e^2\over 4\pi}{1+r^2\over \Box})\partial_\mu\partial_\nu
+{e^2\over
4\pi}{r\over
\Box}(\tilde\partial_\mu\partial_\nu
+\tilde\partial_\nu\partial_\mu)
\label{29}
\en
\noindent
and, consequently,
\bea
(K^{-1})_{\mu\nu}\equiv D_{\mu\nu} & =&{1\over
\Box+m^2}  [g_{\mu\nu}+{\Box+{e^2\over 4\pi}(1+r^2)\over
{e^2\over 4\pi} (a-r^2)} {\partial_\mu\partial_\nu\over
\Box}+ \nonumber\\
& &  +{r\over r^2-a}{1\over
\Box} (\tilde\partial_\mu\partial_\nu+\tilde\partial_\nu\partial_\mu)
 ].
\label{210}
\eea
We have introduced the quantity
\eq
m^2={e^2\over 4\pi}{a (1+a-r^2 )\over a-r^2} ,
\label{211}
\en
\noindent
which is to be interpreted as a dynamically generated mass in the
theory;
$D_{\mu\nu}$ has a pole there $\sim(k^2-m^2+i\epsilon)^{-1}$, with
causal prescription, as usual. We note that
$D_{\mu\nu}$ exhibits also a pole at $k^2=0$.

Eqs.
(\ref{210}) and (\ref{211}) generalize the well--known results of the
vector and chiral Schwinger models.
As a matter of fact, setting first $r=0$ and then $a=0$ we recover for
$m^2$ the value ${e^2\over {4\pi}}$ of the (gauge invariant version of the)
vector Schwinger model.
The kinetic term $K_{\mu\nu}$ becomes a projection operator
\eq
K_{\mu\nu} (a=0, r=0 )= (\Box
+m^2 ) (g_{\mu\nu}-{\partial_\mu\partial_\nu\over
\Box} ),
\label{212}
\en
\noindent
which can only be inverted after imposing a gauge fixing.
In other words the limit $r=0$, $a=0$ in eq.(\ref{210}) is singular, as it
should, as gauge invariance is indeed recovered.

When $r=\pm 1$, we obtain the two equivalent formulations of the chiral
Schwinger model; eq.(\ref{211}) becomes
\eq
m^2={e^2\over 4\pi}{a^2 \over a-1}.
\label{213}
\en

To avoid tachyons, we must require $a>1$.
\noindent
Gauge invariance is definitely lost, and eq.(\ref{210}) becomes
\bea
D_{\mu\nu}&=&{1\over \Box+m^2} [g_{\mu\nu}+{1\over
a-1} ({\pi\over e^2}+{2\over
\Box} )\partial_\mu\partial_\nu \mp \nonumber\\
& & \mp{1\over
a-1}{\tilde\partial_\mu\partial_\nu+\tilde\partial_\nu\partial_\mu\over
\Box} ].
\label{214}
\eea

Going back to the general expression eq.(\ref{211}) we remark that the
condition $m^2>0$, which is necessary to avoid the presence of tachyons
in the theory, allows two ranges:
\bea
1)\qquad\qquad\qquad & a>r^2,&\nonumber\\
2)\qquad\qquad\qquad & 0<a<r^2-1& \qquad\qquad {\rm or} \qquad\qquad
r^2-1<a<0,
\label{216}
\eea
\noindent
for the parameters $(a,r)$. Only the first range has been considered so
far in the literature, to our knowledge.

By taking in eq.(\ref{210}) the residue at the pole $k^2=m^2$, one gets
\eq
Res~D_{\mu\nu}\mid_{k^2=m^2}={1\over m^2}T_{\mu\nu}(k),
\label{217}
\en
\noindent
$T_{\mu\nu}$ being a positive semidefinite degenerate quadratic form in
$k_\mu$, involving the
parameters $(a,r)$. One eigenvalue vanishes, corresponding to a decoupling
of the would--be related excitation, the other is positive and can be
interpreted in both ranges as the presence of a vector particle with
a rest mass given by the positive square root of eq.(\ref{211}) and
positive residue at the pole in agreement with the unitary condition.
\noindent
This state decouples in the limit $a=r^2$.
\noindent
There is also a massless degree of freedom with
\eq
Res~D_{\mu\nu}\mid_{k^2=0}={\pi\over e^2
a(1+a-r^2)} [ (1+r^2 )k_\mu k_\nu-r (\tilde k_\mu
k_\nu+\tilde k_\nu k_\mu ) ]\mid_{k^2=0}.
\label{218}
\en

One can easily realize that again the quadratic form at the numerator is
positive semidefinite for any value of $r$. The poles at $k^2=m^2$ and
$k^2=0$ exhaust the singularities of $D_{\mu\nu}$.

Let us consider the situation in the two ranges of parameters.
The first range does not deserve particular comments at this stage. No
ghost is present at $k^2=0$, as one eigenvalue of the residue matrix
vanishes and the other is positive, corresponding to a ``physical"
excitation.
The second range does entail no news concerning the state with mass $m$.
The situation is different however when considering the pole at $k^2=0$. We
have indeed a negative residue in this case corresponding to a ``ghost"
excitation (particle with a negative probability). The theory can be
accepted only if this excitation can be consistently excluded from a
positive norm Hilbert space of states, which is stable under time
evolution. This can be done, as shown in ref.\cite{Gri94} by means of
a canonical approach.

The bosonic world is rather dull, consisting only of free
excitations; therefore it is worth
considering at this stage the fermionic sector.

We go back to the general expression eq.(\ref{26}) in which fermionic
sources are on. We have now to consider the fermionic propagator in the
field $A_\mu$, which obeys the equation
\eq
 [i\partiall +e {(1-r\gamma^5 )\over 2}\A ] S (x,
y;A_\mu )=\delta^2 (x-y ),
\label{219}
\en
\noindent
with causal boundary conditions.
\noindent
Let us also introduce the free propagator $S_0$
\eq
i\partiall S_0(x)=\delta^2(x)
\label{220}
\en
\noindent
with the solution
\eq
S_0(x)=\int{d^2 k\over (2\pi)^2}{\kbar\over k^2+i\epsilon} e ^{-ikx}={1\over
2\pi}{\gamma_\mu x^\mu\over x^2-i\epsilon}.
\label{221}
\en

If we remember that any vector fields in two dimensional Minkowski space
can be written as a sum of a gradient and a curl part
\eq
A_\mu=\partial_\mu\alpha+\tilde\partial_\mu\beta,
\label{222}
\en
\noindent
the following change of variables in eq.(\ref{24})
\eq
\psi=\exp
 [{ie\over 2} (\alpha+\gamma^5\beta+r\beta+r\alpha\gamma^5 ) ]\chi
\label{223}
\en
\noindent
realizes the decoupling of the fermions, leading to the expression for the
``left" propagator:
\bea
S^L(x-y)&\equiv &\int{\cal D} (A_\mu,\phi ) S^L (x,y;
A_\mu ) \exp\Bigl[i\int d^2z{\cal L}_{eff}(A_\mu,\phi)\Bigr]=\nonumber\\
&=&S_0^L(x-y)Z_L \exp \{-{1\over 4}{(1-r^2)^2\over
a(a+1-r^2)}\ln [\tilde
m^2 (-(x-y)^2+i\epsilon ) ]- \nonumber\\
& & -i\pi{a+1-r^2\over a(a-r^2)} (r-{a\over a+1-r^2} )^2
D (x-y,m ) \},
\label{224}
\eea
\noindent
where $\tilde m={me^\gamma\over 2}$, $D$ is the scalar Feynman
propagator: $D\equiv D_0$, with
\bea
D_{1-\omega} (x, m )&=&- (\lambda^2)^{1-\omega}\int{d^{2\omega}
k\over (2\pi)^{2\omega}}{e^{-ikx}\over
k^2-m^2+i\epsilon}\nonumber\\
&=&{2i\over (4\pi)^{\omega}}
 ({\lambda^2 \sqrt {-x^2}\over 2m} )^{1-\omega}
K_{1-\omega} (m\sqrt{-x^2+i\epsilon} ),
\label{225}
\eea
\noindent
$\gamma$ being the Euler--Mascheroni constant.
For further developments it is useful to consider $2\omega$ dimensions
and to introduce a  mass parameter
$\lambda$ to balance dimensions. $Z_L$ is a (dimensionally
regularized) ultraviolet renormalization constant for the fermion wave
function
\eq
Z_L=\exp [i{\pi(r-1)^2\over a-r^2} D_{1-\omega}(0,m) ].
\label{226}
\en

The ``right" propagator can be obtained from eq.(\ref{224}) simply by
replacing $S_0^L$ with $S_0^R$ and changing the sign of the
parameter $r$.

The Fourier transform in the momentum space of eq.(\ref{224}) cannot
be obtained in closed form; however it exhibits the singularities
related to the thresholds at $p^2=0$ and $p^2=(nm)^2$, $n=1,2,3,...$.

Now we show how to derive the left propagator eq.(\ref{224})
in the path--integral formalism; all the other Green functions can be
obtained in the same way.
\noindent
The first step is to integrate the fermions in eq.(\ref{24}) to give
eq.(\ref{26}) (we put $J_\mu=0$). The change of variables
eq.(\ref{223}) decouples the spinors from $A_\mu$ but has a non trivial
Jacobian ${\cal J}[A_\mu]$
\eq
{\cal J}[A_\mu]=\exp\int d^2x{e^2\over
8\pi}A_\mu \Bigl[\, (1+a )g^{\mu\nu}- (1+r^2 )
{\partial^\mu\partial^\nu\over
\Box}-2r\epsilon^{\alpha\mu}{\partial_\alpha\partial^\nu\over
\Box}\,\Bigr ]A_\nu.
\label{A1}
\en

This result can be obtained, in euclidean space, using $\zeta-$function
technique for functional determinants \cite{Haw77}

\eq
\det [D] = exp [- {d \over ds} \zeta_D(s) |_{s=0}],
\label{g1}
\en
\eq
\zeta_D(s)=\int d^{2n}x Tr  [ K_s(D;x,x) ],
\label{g2}
\en
where $K_s(D;x,x)$ is the kernel of $D^{-s}$, $D$ being
a pseudoelliptic operator \cite{See67}.

The fermionic action is now

\bea
\int &d^2x& \Bigl[i\tilde\chi\partiall\chi+\bar\eta\exp\Bigl(
{ie\over 2} [\alpha+\gamma^5\beta+r\beta+r\alpha\gamma^5 ]\Bigr)\chi
+\nonumber\\
&+&\bar\chi\exp
\Bigl({ie\over 2} [-\alpha+\gamma^5\beta-r\beta+r\alpha\gamma^5 ]\Bigr)
\eta \Bigr ],
\label{A2}
\eea
\noindent
where $\chi$ is a free fermion and $\alpha, \beta$ are linked by
eq.(\ref{222}) to $A_\mu$.
\noindent
The diagonalization of eq.(\ref{A2}) gives the propagator $S_(x,y;A_\mu)$:
\bea
S_(x,y;A_\mu)&=&S^L_0(x-y)\exp \Bigl[i\int
d^2z\, \xi^L_\mu(z;x,y)A^\mu(z) \Bigr]+ \nonumber\\
&+&S^R_0(x-y)\exp \Bigl[i\int
d^2z\xi^R_\mu(z;x,y)A^\mu(z) \Bigr],\nonumber\\
\xi^{L,R}_\mu(z;x,y)&=&{e\over 2}(r\pm
1)(\partial^z_\mu\pm\tilde\partial_\mu^z) [D(z-x)-D(z-y) ],
\label{A3}
\eea
\noindent
where $D(x)$ is the free massless scalar propagator in $d=1+1$ and
$S_0^L$, $S^R_0$ the free left and right fermion propagators.

To obtain the left propagator eq.(\ref{224}) we derive
with respect to $\bar\eta_L$ and $\eta_L$
(the left component of the sources eq.(\ref{25}) and get
\eq
S^L(x,y)=S^L_0(x-y)\int{\cal D}  A_\mu{\cal J}[A_\mu]\exp i\int d^2z [
-{1\over 4}F_{\mu\nu} F^{\mu\nu}(z)+\xi^L_\mu(z;x,y)A^\mu(z) ].
\label{A4}
\en

Using the explicit form of ${\cal J}[A_\mu]$ (eq.(\ref{A1})),
we can write the
path--integral over $A_\mu$ as
\eq
\int{\cal D} A_\mu \exp (i\int d^2 z [\xi^L_\mu A^\mu
+{1\over 2}A_\mu K^{\mu\nu}
A_\nu ] ),
\label{A5}
\en
\noindent
$K^{\mu\nu}$ being defined in eq.(\ref{29}).
The Gaussian integration is trivial and gives
\eq
S_L(x,y)=S^L_0(x-y)\exp (-{1\over 2}\int d^2 zd^2
w\xi^L_\mu(z;x,y)\{K^{-1}\}^{\mu\nu}(z,w)\xi^L_\nu(w;x,y) ).
\label{A6}
\en

The explicit computation of the exponential factor gives the
renormalization constant $Z_L$ and the interaction contribution in
eq.(\ref{224}).

We can now begin the study of the fermionic propagator.
First of all, we notice that for $r=1$ the ``left" fermion is free. The
same  happens to the ``right" fermion when $r=-1$.
Moreover we notice from eq.(\ref{224}) that the long range interaction
completely decouples for $r^2=1$.
As a consequence the interacting fermion (for instance the ``right" one for
$r=1$) asymptotically behaves like a free particle.

In general, at small values of $x^2$, the propagator $S^L$ has the
following behaviour
\eq
S^L\sim_{x^2 \rightarrow 0}C_0 x^+ (-x^2+i\epsilon )^{-1-A}
\label{227}
\en
\noindent
with
\eq
A={1\over 4}{(1-r)^2\over a-r^2}
\label{228}
\en
\noindent
and $C_0$ a suitable constant.

\noindent
We remark that the ultraviolet behaviour of the left fermion propagator
can be directly obtained from the ultraviolet renormalization constant
\eq
\gamma_{\psi_{L}}=\lim_{\omega \rightarrow 1}{1\over 2}
 (\lambda{\partial\over \partial\lambda}\ln Z_L )=-{(1-r)^2\over
4(a-r^2)}
\label{229}
\en
and, of course, it coincides
with the one of the explicit solution eq.(\ref{224}).
It can be obtained from the renormalization group equation in the
ultraviolet limit in which the mass dependent term is disregarded.

For large values of $x^2$ we get instead
\eq
S^L\sim_{x^2 \rightarrow-\infty}~C_\infty
x^+ (-x^2+i\epsilon )^{-1-B}
\label{230}
\en
\noindent
where
\eq
B={1\over 4}{(1-r^2)^2\over a(a+1-r^2)}
\label{231}
\en
\noindent
and $C_\infty$ another constant.

We have shown in \cite{Gri94} that eq.(\ref{230}) exactly coincides with
the behavior of the
fermionic propagator of the massless Thirring model in the
spin-${1\over 2}$ representation, with a coupling constant:
\eq
g={1-r^2\over a}.
\label{tirri1}
\en

We remark that in the first region of the parameter space, where
unitarity holds without subsidiary condition, it happens that
\eq
g>-1,
\label{tirri2}
\en

as required to have a consistent solution for the model \cite{Col75}.

In the first range $(a>r^2)$, both $A$ and $B$ are positive. The
propagator decreases at infinity indicating the possible existence
of asymptotic
states for fermions, which however feel the long range interaction mediated
by the massless excitation which is present in the bosonic spectrum.
The situation in the second range is much more intriguing. Here both $A$
and $B$ are negative. Moreover
\eq
1+B={(2a+1-r^2)^2\over 4a(a+1-r^2)}<0
\label{232}
\en
leading to a propagator which increases when $x^2 \rightarrow -\infty$.
We interpret this phenomenon as a sign of confinement. We recall indeed
that gauge invariance is broken and therefore the fermion propagator is
endowed of a direct physical meaning.
The unphysical massless bosonic excitation, which occurs in this region
of parameters,
produces an anti--screening effect of a long range type.
Nevertheless no asymptotic freedom is expected $(A\not=0)$.

There is another interesting quantity which can be easily discussed
in a path--integral approach.
Let us introduce the scalar fermion composite operator
\eq
\hat{S}(x)=N [\bar\psi(x)\psi(x) ]
\label{e233}
\en
\noindent
where $N$ means the finite part, after divergencies have been
(dimensionally) regularized and renormalized.
\noindent
By repeating standard techniques, it is not difficult to get the expression
\eq
<0\mid T (\hat{S}(x)\hat{S}(0) )\mid 0>=-{Z^{-1}\over
2\pi^2(x^2-i\epsilon)}{\cal K}(x)
\label{234}
\en
\noindent
where
\bea
{\cal K}(x)&=&\exp \{-4i\pi [{a\over
(a-r^2)(a-r^2+1)} (D(x,m)-D_{1-\omega}(0,m) )+  \nonumber\\
&+&  {1-r^2\over
a-r^2+1} (D_{1-\omega}(0,0)-D_{1-\omega}(x,0) ) ] \}
\label{235}
\eea
\noindent
and
\eq
Z=\exp \{4i\pi{r^2\over a-r^2}D_{1-\omega}(0,m) \}.
\label{236}
\en
Dimensional regularization is again understood.

The expressions
for the renormalization constants $Z_L$ and $Z$  we have
just found, will be recovered
in the decompactification limit from the corresponding quantities on
$S^2$.

We end this section by remarking that conformal invariance is
recovered both in the ultraviolet and in the infrared limit, with different
scale coefficients.
\vskip 0.5truecm

\section{Compactification of $R^2$ to $S^2$: non--trivial principal
bundle and problems of globality}

In the previous section we have considered the generalized chiral Schwinger
model on the Minkowski space: now we want to study its euclidean version
on a
compact riemannian manifold, namely on the two--dimensional sphere
$S^{2}$.

There are many reasons for this investigation: it is well
known that in the (vector) Schwinger model gauge field configurations
with non--trivial topology (winding number different from zero) and
zero modes of the Dirac operator play an important role in order to
identify the vacuum structure of the theory. More precisely, we can
consider Q.E.D. on $S^2$ which, in the limit of the radius $R$ going to
infinity, becomes Q.E.D. in the euclidean two--dimensions. One may say
that the two--dimensional plane is compactified to $S^2$. This kind of
compactification is particularly suited for studying the mentioned
problems.
Because of the non--trivial topology of $S^2$, the gauge fields fall
into classes characterized by the winding number $n$, defined as:
\eq
n={e\over 2\pi}\int_{S^2} d^{2}x\, F_{01}(x),
\label{s1}
\en

which is an integer. $A_{\mu}$ belongs to a
non--trivial principal bundle over $S^2$. The number of zero modes of
the Dirac operator, linked to $A_{\mu}$, turns out to be equal to $|n|$.
Thus to neglect the zero modes is equivalent to neglecting all
non--trivial topological sectors and leads to an incorrect result even
in the limit $R\rightarrow\infty$. In particular it entails the
vanishing of the fermionic condensate
$<\bar{\psi}\psi>$, in disagreement with an operatorial analysis.

We show that in order to define the
determinant of the Dirac--Weyl operator an external fixed
background connection
must be introduced: the determinant shall depend on it. This is a new
feature that appears in an anomalous gauge theory: as the coboundary
terms become relevant, even the background connection plays an important
role. The quantum theory has indeed an intrinsic arbitrariness which
depends this time not just on a parameter, but on a field configuration
(the choice of the coboundary).

We construct the Green functions generating functional for finite
$R$, defining the theory on the two--sphere. As an application,
we shall eventually compute
the fermionic propagator and the value of
the condensate $<\bar{\psi}\psi>$ in the limit $R\rightarrow\infty$.

Before examining the generalized model we are interested in, some
geometrical considerations are in order and the simpler cases of the
vector and chiral models have to be first thoroughly discussed.
\vskip 0.5truecm

Our first step is to give a geometrical description of $U(1)$--valued
one--forms on the two--sphere, using angular and stereographical
coordinates for $S^2$ (of radius $R$).

We can parametrize the two--sphere by angular coordinates:
\bea
\theta,\varphi\quad\quad\quad
&0\leq\theta<\pi&\nonumber\\
&0\leq\varphi<2\pi&
\label{s2}
\eea

or by the use of the stereographical projection (the south pole is
identified with the $\infty$):
\bea
\hat{x}_{1},\hat{x}_{2}
\quad\quad\quad
\hat{x}_{1}&=&2R\tan{\theta\over 2}\cos\varphi, \nonumber\\
\hat{x}_{2}&=&2R\tan{\theta\over 2}\sin\varphi.
\label{s3}
\eea

The natural metric is:
\bea
g_{\th\th}&=&R^{2}\nonumber\\
g_{\mu\nu}=\quad g_{\varphi\varphi}&=&R^2 \sin^{2}\th\nonumber\\
g_{\th\varphi}&=&0.
\label{s4}
\eea

An orthonormal basis for the tangent space can be chosen:
\bea
e_{1}&=&{1\over R}{\partial\over\partial\th},\nonumber\\
e_{2}&=&{1\over R\sin\th}{\partial\over\partial\varphi}.
\label{s5}
\eea

A basis for the one--forms is obviously:
\bea
\hat{e}_{1}&=&R\,d\th,\nonumber\\
\hat{e}_{2}&=&R\,\sin\th\,d\varphi.
\label{s6}
\eea

We define the $U(1)$--valued one--form:
\eq
A=A_{\th}(R\,d\th)+A_{\varphi}(R\,\sin\th\,d\varphi).
\label{s7}
\en

The corresponding object in stereographical coordinates is:
\bea
A&=&A_{\th}R\Bigl[{d\th\over d\hat{x}_{1}}d\hat{x}_{1}+{d\th\over
d\hat{x}_{2}}d\hat{x}_{2}\Bigr]+\nonumber\\
&+&A_{\varphi} R\sin\th\Bigl[ {d\varphi\over d\hat{x}_{1}}d\hat{x}_{1}+
{d\varphi\over d\hat{x}_{2}}d\hat{x}_{2}\Bigr].
\label{s8}
\eea

Now
\bea
\th&=&2\arctan\Bigl[{1\over
2R}\sqrt{\hat{x}_{1}^{2}+\hat{x}_{2}^{2}}\Bigr],\nonumber\\
\varphi&=&\arctan\Bigl[{\hat{x}_{2}\over\hat{x}_{1}}\Bigr],
\label{s9}
\eea

leading to
\bea
A&=&\hat{A}_{1}d\hat{x}_{1}+\hat{A}_{2}d\hat{x}_{2},\nonumber\\
\hat{A}_{1}&=&{1\over 1+{\hat{x}^{2}\over 4R^2}}\Bigl[
A_{\th}{\hat{x}_{1}\over \sqrt{\hat{x}^{2}}}-A_{\varphi}{\hat{x}_{2}\over
\sqrt{\hat{x}^{2}}}\Bigr],\nonumber\\
\hat{A}_{2}&=&{1\over 1+{\hat{x}^{2}\over 4R^2}}\Bigl[
A_{\th}{\hat{x}_{2}\over \sqrt{\hat{x}^{2}}}+A_{\varphi}{\hat{x}_{1}\over
\sqrt{\hat{x}^{2}}}\Bigr].
\label{s10}
\eea

The stereographical
projection establishes a one to one correspondence between the points of
the plane and the points of the sphere except the $\infty$ that is
identified with the south pole. Let us consider (in stereo--coordinates)
the connection:
\eq
\hat{A}^{(n)}_{\mu}=-{n\over e}\epsilon_{\mu\nu}{\hat{x}_{\nu}\over
4R^2+\hat{x}^{2}};
\label{s11}
\en

in angular coordinates:
\bea
A^{(n)}_{\th}&=&0,\nonumber\\
A^{(n)}_{\varphi}&=&{n\over 2eR}\tan{\th\over 2},
\label{s12}
\eea

namely
\eq
A^{(n)}={n\over 2e}\tan{\th\over 2}(\sin\th\,d\varphi).
\label{s13}
\en

We observe that the one--form $(\sin\th\,d\varphi)$ has a global meaning
while a singularity arises in the $\varphi$--component: in order to
understand if this singularity is meaningful or it is only an artifact of
our coordinate system (we stress that at least two patches are needed
to describe a sphere and therefore a singularity might be a spurious effect)
, we study the situation in another patch.

The previous result can be rewritten as:
\eq
A^{(n)}={n\over 2e}(1-\cos\th)d\varphi,
\label{s14}
\en

being regular in a region containing $\th=0$ and excluding $\th=\pi$:
\eq
0\leq\th<\pi.
\label{s15}
\en

Now we consider stereographical coordinates derived from a north pole
projection: it is a simple exercise to show that the relation between
the ``northern'' and the ``southern'' coordinates is
\eq
\hat{x}_{S}={4R^2\over \hat{x}_{N}^{2}}(-\hat{x}_{N\,1},\hat{x}_{N\,2})
\label{s16}
\en

and that the connection eq.(\ref{s11}) has the same form. We can repeat
all the calculations finding the expression of $A^{'}$:
\eq
{A^{(n)}}^{'}={n\over 2e}(1+\cos\th)d\varphi,
\label{s17}
\en

that, this time, is well defined on $0<\th\leq\pi$ (the south pole,
$\th=\pi$ is safe being mapped in a finite plane coordinate).

We immediately notice that in the intersection of the patches,
$0<\th<\pi$, $A^{(n)}$ and ${A^{(n)}}^{'}$ do not coincide:
we have two different
expressions for $A$, that cannot be globally defined on the sphere.
Nevertheless in the patch intersection:
\bea
A^{(n)}-{A^{(n)}}^{'}&=&{n\over e}d\varphi={1\over e}i\,g^{-1}dg,\nonumber\\
g&=&\exp[-in\varphi],
\label{s18}
\eea

$g$ being a map from the intersection region to $U(1)$
(notice that this is possible only
if $n$ is an integer); ${A^{(n)}}^{'}$ differs from $A^{(n)}$ by a
gauge transformation.
Gauge invariant objects possess a global definition on $S^2$: $A^{(n)}$ belongs
to
a non--trivial principal bundle on $S^2$. Then:
\bea
d\N&=&{n\over 2e}\sin\th\,d\th\,d\varphi,\nonumber\\
{e\over 2\pi}\int d\N &=&n,
\label{s19}
\eea

and $\N$ carries the non--trivial winding number $n$.

All the connections on the plane that can be considered as derived by a
process of stereographic projection, carry integer winding number and
belong, on the sphere, to a $U(1)$--bundle characterized by the same integer.
In general we can represent any connection as
\eq
A_{\mu}=\N_{\mu}+a_{\mu},
\label{s20}
\en

where obviously:
\[{e\over
2\pi}\int_{S^2}d^{2}x\,(\partial_{0}a_{1}-\partial_{1}a_{0})=0.\]
All the topological charge is carried by $\N_{\mu}$ and $a_{\mu}$ admits
a global representation on $S^2$.

\vskip 0.5truecm
\section{Dirac and Dirac--Weyl operators on $S^2$}

Let us take for the moment angular coordinates: to build the Dirac
operator we need the zwei--bein related to the metric eq.(\ref{s4}):
\bea
e^{1}_{\th}&=&R\cos\varphi,\nonumber\\
e^{1}_{\varphi}&=&-R\sin\varphi\sin\th,\nonumber\\
e^{2}_{\th}&=&R\sin\varphi,\nonumber\\
e^{2}_{\varphi}&=&R\cos\varphi\sin\th.
\label{s21}
\eea

The Dirac operator is (we choose $A_{\mu}=\N_{\mu}$):
\eq
D=i\gamma_{a}e^{\mu}_{a}\Bigl[\partial_{\mu}+{1\over
4}\gamma_{c}\gamma_{d}\omega_{\mu cd}+ie\N_{\mu}\Bigr],
\label{s22}
\en

$\omega_{\mu cd}$ being the usual spin--connection. We give the only non
zero component (we are in the ``southern'' patch $0\leq\th<\pi$):
\eq
\omega_{\varphi 12}=1-\cos\th.
\label{s23}
\en

The Dirac operator is:
\eq
D=\left( \begin{array}{cc}  0  &  D_{12} \\
                                      D_{21} &  0  \end{array}
                                                              \right)
\label{s24}
\en

where
\bea
D_{12}&=&i\exp(-i\varphi)\Bigl[\partial_{\th}-{i\over
\sin\th}\partial_{\varphi}-{1-n\over 2}{1-\cos\th\over
\sin\th}\Bigr],\nonumber\\
D_{21}&=&i\exp(i\varphi)\Bigl[\partial_{\th}+{i\over
\sin\th}\partial_{\varphi}-{1+n\over 2}{1-\cos\th\over
\sin\th}\Bigr].
\label{s25}
\eea

It is very simple to derive the explicit expression of this operator in
the second patch:
\eq
D^{'}=\left( \begin{array}{cc}  0  &  D^{'}_{12} \\
                                      D^{'}_{21} &  0  \end{array}
                                                              \right),
\label{s26}
\en

\bea
D^{'}_{12}&=&i\exp(i\varphi)\Bigl[\partial_{\th}-{i\over
\sin\th}\partial_{\varphi}+{1-n\over 2}{1+\cos\th\over
\sin\th}\Bigr],\nonumber\\
D^{'}_{21}&=&i\exp(-i\varphi)\Bigl[\partial_{\th}+{i\over
\sin\th}\partial_{\varphi}+{1+n\over 2}{1+\cos\th\over
\sin\th}\Bigr].
\label{s27}
\eea

In the intersection of the patches the two operators are related by a
unitary transformation:
\[D^{'}=U^{-1}D\,U,\]
\eq
U=\left( \begin{array}{cc}  \exp[-i(n+1)\varphi]  &  0  \\
                                     0   &  \exp[i(n+1)\varphi]  \end{array}
                                                              \right).
\label{s28}
\en

$D$ maps a globally defined Dirac field into a new one; the eigenvalue
equation, that is essential to obtain the Dirac determinant,
\[D\psi=E\,\psi,\]
has therefore a well defined meaning with all the eigenvalues $E$'s being
invariant under gauge transformations and local frame rotations.

The situation for a Dirac--Weyl operator
\eq
D=i\gamma_{a}e^{\mu}_{a}\Bigl[\partial_{\mu}+{1\over
4}\gamma_{c}\gamma_{d}\omega_{\mu cd}+ie({1+\gamma_{5}\over 2})\N_{\mu}\Bigr],
\label{s29}
\en

is completely different owing to the relation:
\[D^{'}=U_{1}D\,U_{2},\]
\eq
U_{1}=\left( \begin{array}{cc}  \exp[-i\varphi]  &  0  \\
                                     0   &  \exp[i(n+1)\varphi]  \end{array}
                                                              \right),
\label{s30}
\en
\eq
U_{2}=\left( \begin{array}{cc}  \exp[-i(n+1)\varphi]  &  0  \\
                                     0   &  \exp[i\varphi]  \end{array}
                                                              \right).
\label{s31}
\en

The eigenvalue equation in this case has no global meaning: for a
generic $A_{\mu}$ of winding number $n$ (see eq.(\ref{s20}))
the situation does not change.

It is well known that, in presence of a non--trivial fiber bundle,
globality considerations force the dependence of the anomaly on a
fixed background gauge connection \cite{Stora85}, \cite{rfv1}, \cite{rfbar}:
the vertex functional is assumed to involve both a dynamical gauge field $A$
and an ``external'' one $A_{0}$. From the geometrical point of view this
property
is very simple to be understood: the transgression formula \cite{rfq}
relies on the
fact that a symmetric polynomial on the Lie algebra, invariant under the
adjoint action of the group, usually denoted as $P(F^{n})$, is an exact
form defined on the whole principal bundle while its projection,
considered as a form on the base manifold, is only closed:
\eq
P(F^{n})-P(F_{0}^{n})=d\omega^{0}_{2n-1}(A,A_{0}).
\label{s32}
\en

The anomaly, being derived from $\omega^{0}_{2n-1}(A,A_{0})$,
depends on $A_{0}$: we recall that in this approach \cite{Stora85}
$A_{0}$ does not transform under the B.R.S.T. action, that defines the
cohomological problem, and its introduction makes the solution globally
defined. A change of the background connection
reflects itself in a change of the coboundaries of the cohomological
solution \cite{Stora85}: in this sense the choice of $A_{0}$ does not change
the anomaly because the cohomology class remains the same.

We are therefore induced to solve the problem of globality of the
Dirac--Weyl operator in a similar way:
we introduce a fixed background connection, belonging to the same bundle,
in order to recover the transformation property eq.(\ref{s28}) in passing
from a patch to another:
\eq
D=i\gamma_{a}e^{\mu}_{a}\Bigl[\partial_{\mu}+{1\over
4}\gamma_{c}\gamma_{d}\omega_{\mu cd}+ie\,({1+\gamma_{5}\over 2})A_{\mu}+
ie\,({1-\gamma_{5}\over 2})A^{0}_{\mu}\Bigr].
\label{s33}
\en

It is rather clear that
\[D^{'}=U^{-1}D\,U\]
and hence the global meaning of the Dirac--Weyl determinant is safe
\cite{rfbar}.

It is an exercise to compute the gauge anomaly from the operator
eq.(\ref{s33}): one can use the $\zeta$-function technology to recover
the infinitesimal variation of the $D$--determinant, that coincides with
the result of \cite{Stora85}. Different choices of $A_{0}$
reflect themselves into different representatives of the cohomology class:
a change of $A_0$ changes the local terms of the determinant. This is a
new feature we find in studying an anomalous model on a compact surface:
we stress again that
an anomalous theory strictly depends on the choice of the coboundary so
that the quantum theory looks sensitive to this background connection.
In the next section we will try to understand this dependence, and to
arrive to a reasonable definition of a chiral gauge theory on the
sphere.

\vskip 0.5truecm
\section{The chiral gauge theory on $S^2$}
The gauge fields on $S^2$ fall into classes characterized by
the topological charge $n$:
\bea
n&=&{e\over
2\pi}\int_{S^2}d^{2}x\sqrt{g}\epsilon^{\mu\nu}F_{\mu\nu}\nonumber\\
\epsilon^{01}&=&{1\over \sqrt{g}}=-\epsilon^{10}.
\label{s34}
\eea

Let us consider the field $\N_{\mu}$, defined in stereographical
coordinates by eq.(\ref{s11}): $F^{(n)}_{\mu\nu}$ turns out to be
\eq
F^{(n)}_{\mu\nu}={n\over 2eR^{2}}\epsilon_{\mu\nu}
\label{s35}
\en

and satisfies the equation of motion ($D_{\mu}$ is the
covariant derivative with respect to the usual Levi--Civita connection)
\eq
D_{\mu}F^{(n)\mu\nu}=0.
\label{s36}
\en

Obviously eq.(\ref{s36}) has a global meaning due its gauge invariance,
while $\N_{\mu}$ does not possess a global expression. In the same way any
field
of the type
\eq
\tilde{A}^{(n)}_{\mu}= \N_{\mu}+{1\over ie}u\partial_{\mu}u^{-1},\nonumber\\
\label{s37}
\en

where $u$ is a $U(1)$--valued map, is a solution of eq.(\ref{s36}). In
particular it happens that:
\eq
F^{(n)}_{01}={n\over 2eR^2}\sqrt{g}.
\label{s38}
\en

Now let us suppose that $A_\mu$ is any gauge potential with topological charge
$n$: a field $\phi$ can be defined through
\bea
-\Delta\phi&=&{F^{(n)}_{01}\over \sqrt{g}}-{n\over 2eR^2},\nonumber\\
\int_{S^2}d^{2}x\sqrt{g}\,\phi &=&0,
\label{s39}
\eea

 ${F^{(n)}_{01}\over \sqrt{g}}$ being a scalar field on $S^2$; the Laplace
--Beltrami operator is:
\eq
\Delta={1\over \sqrt{g}}\partial_{\mu}\sqrt{g}g^{\mu\nu}\partial_{\nu}.
\label{s40}
\en

If we expand the function ${F^{(n)}_{01}\over \sqrt{g}}$ in a complete
orthonormal set of eigenfunctions of $\Delta$, the term ${n\over
2eR^2}$ corresponds to its zero mode. Hence the
function ${F^{(n)}_{01}\over \sqrt{g}}-{n\over
2eR^2}$ has no projection on the zero mode and the
laplacian can be inverted to obtain $\phi$, that is gauge invariant and
scalar under diffeomorphism. The most general form of a generic
$A_{\mu}$ is thereby:
\eq
A_{\mu}=\N_{\mu}+\epsilon_{\mu\nu}g^{\nu\rho}\partial_{\rho}\phi+{i\over
e}h\partial_{\mu}h^{-1},
\label{s41}
\en

$h\in U(1)$. The winding number is carried by $\N$, $\phi$ and $h$
describing the topologically trivial part.

Now we define the chiral Schwinger model on the sphere by:
\bea
S^{(n)}_{Class.}&=&\inta {1\over
4}F^{(n)}_{\mu\nu}F^{(n)}_{\rho\lambda}g^{\mu\rho}g^{\nu\lambda}+ \nonumber\\
&+&\bar{\psi}\,\gamma_{a}\,e^{\mu}_{a}\Bigl[\,iD_{\mu}+e\N_{\mu}+
e({1+\gamma_{5}\over
2})a_{\mu}\,\Bigr]\,\psi\nonumber\\
F^{(n)}_{\mu\nu}&=&\partial_{\mu}[\N_{\nu}+a_{\nu}]
-\partial_{\nu}[\N_{\mu}+a_{\mu}],\nonumber\\
a_{\mu}&=&\epsilon_{\mu\nu}g^{\nu\rho}\partial_{\rho}\phi+{i\over
e}h\partial_{\mu}h^{-1}.
\label{s42}
\eea

$S^{(n)}_{Class.}$ is the action on the $n$--topological sector: now
we define any
expectation value of quantum operators $O(\bar{\psi},\psi,A)$ as:

\bea
<O(\bar{\psi},\psi,A)>&=&{\cal Z}_0^{-1}\sum^{+\infty}_{n=-\infty}\int{\cal
D}a_{\mu}
{\cal D}\bar{\psi}{\cal D}\psi
\,O(\bar{\psi},\psi.A)\exp\Bigl[-S^{(n)}_{Class.}\Bigr],
\nonumber\\
{\cal Z}_0&=&\sum^{+\infty}_{n=-\infty}\int{\cal D}a_{\mu}
{\cal D}\bar{\psi}{\cal D}\psi \exp-\Bigl[S^{(n)}_{Class.}\Bigr].
\label{s43}
\eea

In so doing we represent the $A_{\mu}$ connection as the
sum of a classical instantonic solution ($\N_{\mu}$) and a quantum
fluctuation ($a_{\mu}$) and we choose the fixed background connection
$A^{0}_{\mu}$ to be equal to $\N_{\mu}$.
We notice that
$S^{(n)}_{Class.}$ does not change under the transformation:
\[
A_{\mu}^{(n)}\rightarrow\tilde{A}_{\mu}^{(n)},
\]
\[
\bar{\psi}\rightarrow\bar{\psi}u^{-1},
\]
\[
\psi\rightarrow u\psi.
\]
Quantum theory is unchanged too, since Jacobian is unity ( as one could
check using $\zeta$--function regularization) for this transformation.

The quantum fluctuation $a_{\mu}$ couples chirally to the spinor field:
no problem of globality arises, having $a_{\mu}$ a global expression.
Our definition is the most natural generalization of the flat case
(or $n=0$) where:
\eq
S^{(0)}_{Class.}=\int d^{2}x\Bigl[{1\over 4}F^{\mu\nu}F_{\mu\nu}+
\bar{\psi}\,i\gamma^{\mu}[\partial_{\mu}+ie\,({1+\gamma_{5}\over
2})A_{\mu}]\psi\Bigr].
\en

The field $A_{\mu}$ is the candidate to represent quantum fluctuations
whereas one chooses, in absence of spinors, as solution
of the equation of motion $A^{(0)}_{\mu}=0$. The fluctuation couples
chirally to $\psi$.

We argue that in a topological sector the vacuum is described by the
classical $\N_{\mu}$ solution, and both components of the spinor have
to interact with it.
But one could also consider eq.(\ref{s43})
as the very definition of our model.

It is rather simple to show that:
\bea
\inta {1\over
4}F^{(n)}_{\mu\nu}F^{(n)}_{\rho\lambda}g^{\mu\rho}g^{\nu\lambda}&=&
{\pi n^{2}\over 2eR^{2}}+\inta {1\over
4}f_{\mu\nu}f_{\rho\lambda}g^{\mu\rho}g^{\nu\lambda},\nonumber\\
f_{\mu\nu}&=&\partial_{\mu}a_{\nu}-\partial_{\nu}a_{\mu}.
\label{s44}
\eea

No coupling between  quantum fluctuation $a_{\mu}$ and classical
background $\N_{\mu}$ arises, due to eq.(\ref{s36}). To get
the action for the generalized chiral Schwinger model we have only to
couple $a_{\mu}$ with:
\[e({1-r\gamma_{5}\over 2}).\]

\vskip 0.5truecm

\section{The Green's function generating functional of the generalized
chiral Schwinger model on $S^{2}$}

In this section we obtain the Green's function generating functional of the
model. The action eq.(\ref{s42}) is again quadratic in fermion fields,
therefore the fermionic integration can be performed for many important
operators. Before performing this integration, let us note that it is more
convenient to use a dimensionless operator in the action. This can be
achieved by setting:
\bea
\psi_{A}&=&{\psi\over \sqrt{R}},\nonumber\\
\bar{\psi}_{A}&=&{\bar{\psi}\over \sqrt{R}},\nonumber\\
\hat{D}&=&R\,D.
\label{s45}
\eea

The operator $\hat{D}$ is not hermitian and possesses a non-trivial
kernel; the result of the fermionic integration depends crucially on the
number of zero modes. In order to work with an hermitian operator (that
in $S^2$ admits a complete set of eigenstates and eigenvalues) and to
bypass problems with the Berezin integration, we define the Green's
function eq.(\ref{s43}) through a process of analytic continuation on a
parameter that we call $\lambda$, a trick essentially due to Andrianov
and Bonora \cite{rfg}:
\bea
<O(\bar{\psi},\psi,A)>&=&\lim_{\lb\rightarrow ir}{\cal Z}_0^{-1}(\lb)
\sum^{+\infty}_{n=-\infty}\int{\cal D}a_{\mu}
{\cal D}\bar{\psi}_{A}{\cal D}\psi_{A} \,
O(\sqrt{R}\bar{\psi}_{A},\sqrt{R}\psi_{A},A)\nonumber\\
& &\exp\Bigl[-{\pi n^{2}\over 2e^{2}R^{2}}-\inta {1\over
4}f_{\mu\nu}f^{\mu\nu}\Bigr]\nonumber\\
& &\exp \Bigl[-\inta R\,\bar{\psi}_{A}\gamma_{a} e^{\mu}_{a}[\,
iD_{\mu}+e\N_{\mu}+e({1+i\lb\gamma_{5}\over 2})a_{\mu}]\psi_{A}\Bigl].
\label{s46}
\eea

We use the properties of the two--dimensional Dirac algebra to get:
\eq
\gamma_{a} e^{\mu}_{a}({1+i\lb\gamma_{5}\over 2})a_{\mu}=
\gamma_{a} e^{\mu}_{a}{1\over 2}(g_{\mu\nu}+\lb
\epsilon_{\mu\nu})a^{\nu}.
\label{s47}
\en

Redefining the (topologically trivial) fluctuation $\hat{a}_{\mu}$:
\bea
{1\over 2}(g_{\mu\nu}+\lb
\epsilon_{\mu\nu})a^{\nu}&=&\hat{a}_{\mu},\nonumber\\
{2 \over 1+\lb^{2}}(g_{\mu\nu}-\lb
\epsilon_{\mu\nu})\hat{a}^{\nu}&=&a_{\mu},
\label{s48}
\eea

we can change variables of integration from $a_{\mu}$ to $\hat{a}_\mu$
(up to an overall constant Jacobian that disappears in eq.(\ref{s46}))
and write:

\bea
& &<O(\bar{\psi},\psi,A)>=\lim_{\lb\rightarrow ir}{\cal Z}_0^{-1}(\lb)
\sum^{+\infty}_{n=-\infty}\exp-[{\pi n^{2}\over 2e^{2}R^{2}}]
\int{\cal D}\hat{a}_{\mu}
{\cal D}\bar{\psi}_{A}{\cal D}\psi_{A}
O(\sqrt{R}\bar{\psi}_{A},\sqrt{R}\psi_{A},A[\,\hat{a}\,])
\nonumber\\
& & \exp \Bigl[\inta{2\over 1+\lb^{2}}\,\hat{a}_{\mu}\Bigl[\,
(g^{\mu\nu}\Delta-D^{\mu}D^{\nu}-R^{\mu\nu})+\lb^{2}D^{\mu}D^{\nu}+
2\lb\epsilon^{\rho\nu}D_{\rho}D^{\mu}\Bigr]\hat{a}_{\nu}\Bigr]\nonumber\\
& & \exp\Bigl[-\inta
R\,\bar{\psi}_{A}\gamma_{a}e^{\mu}_{a}[\,iD_{\mu}+e\N_{\mu}+e
\hat{a}_{\mu}\,]\psi_{A}\Bigr],
\label{s49}
\eea

$R_{\mu\nu}$ being the Ricci tensor on $S^2$.
In this way we have a fermionic integration linked to the Dirac operator (no
chiral couplings); the price we have paid is to work with a more
complicated action for the gauge fluctuation. We remark that the fixed
background connection was not touched by our definition; the
geometrical structure of the theory is unchanged.

Let us study some properties of the operator
\eq
\hat{D}^{(n)}=R\,\gamma_{a}e^{\mu}_{a}[\,iD_{\mu}+e\N_{\mu}+e
\hat{a}_{\mu}\,];
\label{s50}
\en

it can be proved that to every eigenfunction
$\Phi_{i}$ of the operator $\hat{D}^{(n)}$ with a non-zero
eigenvalue $E_{i}$ another eigenfunction $\Phi_{-i}=
\gamma_{5}\Phi_{i}$ corresponds, with eigenvalue $E_i=-E_{-i}$. Furthermore,
the zero modes $\chi^{(n)}_{m}$ have definite chirality i.e.
\[ \gamma_{5}\chi^{(n)}_{m}=\pm \chi^{(n)}_{m},\]
and the number of zero modes is $n=n_{+}+n_{-}$, $n_{+}$ corresponding
to positive chirality and $n_{-}$ to the negative one \cite{ni77}:
\bea
&n_{+}=0&\quad\quad n_{-}=|n| \quad\quad n\geq 0\nonumber,\\
&n_{+}=|n|&\quad\quad n_{-}=0  \quad\quad n\leq 0.
\label{s51}
\eea

We can now compute the partition function by performing the quadratic
fermionic integration:
\eq
{\cal Z}_0(\lambda)=\sum^{+\infty}_{n=-\infty}\int{\cal D}\hat{a}_{\mu}
\exp \Bigl[\, -{\pi n^2 \over
2e^{2}R^{2}}-S_{Bos.}(\lb)\,\Bigr]\det[\hat{D}^{(n)}],
\label{s52}
\en

$S_{Bos.}(\lb)$ being the bosonic part of the action:
\eq
S_{Bos.}(\lb)=-{2\over 1+\lb^{2}}\,\inta\hat{a}_{\mu}\Bigl[\,
(g^{\mu\nu}\Delta-D^{\mu}D^{\nu}-R^{\mu\nu})+\lb^{2}D^{\mu}D^{\nu}+
2\lb\epsilon^{\rho\nu}D_{\rho}D^{\mu}\Bigr]\hat{a}_{\nu}.
\label{ss1}
\en

The presence of the zero modes for $n\neq 0$ leads to a vanishing
contribution of the topological sectors to the partition function:
the determinant, defined as the product of the eigenvalues, is zero in
this case,
\eq
{\cal Z}_0(\lb)=\int{\cal D}\hat{a}_{\mu}
\exp \Bigl[\,-S_{Bos.}(\lb)\,\Bigr]\det[\hat{D}^{(0)}].
\label{s53}
\en

It follows that any operator of the type $O(A)$ takes contribution only
from the $n=0$ sector; on the other hand the correlation functions involving
fermions feel the presence of topological charged configurations:
they manifest themselves through parity violating amplitudes \cite{ho79}.
We will not consider in the following the general problem of mixed
correlation functions, being satisfied with understanding the pure
fermionic and bosonic sectors. To this aim we introduce sources to build the
generating functional:
\bea
{\cal Z}[J_{\mu},\eta,\bar{\eta}]&=&{\cal Z}_0^{-1}
\sum^{+\infty}_{n=-\infty}\int{\cal D}\hat{a}_{\mu}
\exp \Bigl[\, -{\pi n^2 \over
2e^{2}R^{2}}-S_{Bos.}(\lb)-J^{\mu}{2 \over 1+\lb^{2}}(g_{\mu\nu}-\lb
\epsilon_{\mu\nu})\hat{a}^{\nu}\,\Bigr]\nonumber\\
& &\int{\cal D}\bar{\psi}_{A}{\cal D}{\psi}_{A}\exp\Bigl[-\inta
\bar{\psi}_{A}\hat{D}^{(n)}\psi_{A}+\sqrt{R}\bar{\eta}\psi_{A}
+\sqrt{R}\bar{\psi}_{A}\eta\Bigr].
\label{s54}
\eea

Let $\Phi^{(n)}_{i}$ be any eigenstate of $\hat{D}^{(n)}$ corresponding
to a non vanishing eigenvalue: we can
construct the kernel
\eq
\sum_{i\neq 0}{\Phi^{(n)}_{i}(x)\Phi^{(n)\dagger}_{i}(y)\over E_{i}}=
S^{(n)}(x,y),
\label{s55}
\en

that satisfies
\eq
\hat{D}^{(n)}_{x} S^{(n)}(x,y)={\delta^{2}(x,y) \over
\sqrt{g}}-\sum_{m=1}^{n}\chi^{(n)}_{m}(x)\chi^{(n)\dagger}_{m}(y).
\label{s56}
\en

After the translation
\bea
\psi_{A}&=&\psi_{A}^{'}+\int\ d^{2}y
\sqrt{g}\sqrt{R}\,S^{(n)}(x,y)\eta(y),\nonumber\\
\bar{\psi}_{A}&=&\bar{\psi}_{A}^{'}+\inta
\sqrt{R}\,\bar{\eta}(y)S^{(n)}(x,y),
\label{s57}
\eea

the Berezin integration gives:
\bea
{\cal Z}[J_{\mu},\eta,\bar{\eta}]&=&{\cal Z}_0^{-1}
\sum_{n=-\infty}^{+\infty}\int {\cal
D}\hat{a}_{\mu}\exp[-{\pi n^2 \over 2e^2 R^2}-
S_{Bos.}(\lb)-J^{\mu}{2 \over 1+\lb^{2}}(g_{\mu\nu}-\lb
\epsilon_{\mu\nu})\hat{a}^{\nu}]\nonumber\\
& &
\exp[\int
d^{2}x\sqrt{g}\,d^{2}y\sqrt{g}R\,\bar{\eta}(x)S^{(n)}(x,y)\eta(y)]\nonumber\\
& &{\det}^{'}[\hat{D}^{(n)}]\Pi_{m=1}^{n}\Bigl[\inta
\sqrt{R}\chi^{(n)\dagger}_{m}\eta\Bigr]\Bigl[\inta\sqrt{R}\bar{\eta}\chi^{(n)}
_{m}\Bigr];
\label{s58}
\eea

${\det}^{'}[\hat{D}^{(n)}]$ is the (regularized) product of the
non--vanishing eigenvalues. One immediately realizes that the correlation
functions different from zero are of the type:
\eq
<\bar{\psi}(x_{1})\bar{\psi}(x_{2})...\bar{\psi}(x_{N}){\psi}(x_{N+1}){\psi}
(x_{N+2})...\psi(x_{2N})>
\label{s59}
\en

and, at fixed $N$, all the sectors $|n|\leq N$ contribute. At this
point we use the representation for $\hat{a}_{\mu}$:
\eq
\hat{a}_{\mu}=\epsilon_{\mu\nu}g^{\nu\rho}\partial_{\rho}\phi+{1\over
ie}h\partial_{\mu}h^{-1},
\label{s60}
\en

to obtain
\bea
\hat{D}^{(n)}&=&h\exp[e\,\phi\,\gamma_{5}]\hat{D}^{(n)}_{0}\exp[e\,\phi\,
\gamma_{5}]
h^{-1},\nonumber\\
\hat{D}^{(n)}_{0}&=&R\gamma_{a}e^{\mu}_{a}[iD_{\mu}+e\N_{\mu}].
\label{s61}
\eea

The zero modes of $\hat{D}^{(n)}$ are related to the ones of
$\hat{D}^{(n)}_{0}$:
\eq
\chi^{(n)}_{m}=h\exp[-e\phi\gamma_{5}]\sum_{j=1}^{n}B_{mj}\chi^{(n)}_{0j},
\label{s61bis}
\en

where we have used a $n\times n$ matrix $B$ to have an orthonormal basis
for the null space of $\hat{D}^{(n)}$ ($\chi^{(n)}_{0j}$ are chosen to be
orthonormal). The role of the matrix $B$ is discussed in Appendix A.
In eq.(\ref{s58}) one can prove that:
\bea
& &
\exp\Bigl[\int d^{2}x\sqrt{g}d^{2}y\sqrt{g}R\bar{\eta}(x)S^{(n)}(x,y)\eta(y)
\Bigr]\nonumber\\
& &\Pi_{m=1}^{n}\Bigl[\inta
\sqrt{R}\chi^{(n)\dagger}_{m}\eta\Bigr]
\Bigl[\inta\sqrt{R}\bar{\eta}\chi^{(n)}_{m}\Bigr]=\nonumber\\
& &\exp\Bigl[\int
d^{2}x\sqrt{g}d^{2}y\sqrt{g}R\bar{\eta}^{'}(x)S^{(n)}_{0}(x,y)\eta^{'}(y)\Bigr]
\nonumber\\
& &\Pi_{m=1}^{n}\Bigl[\inta
\sqrt{R}\chi^{(n)\dagger}_{0m}\eta^{'}][\inta\sqrt{R}\bar{\eta}^{'}
\chi^{(n)}_{0m}\Bigr]{|\det B|}^{2},
\label{s62}
\eea

where:
\bea
S^{(n)}_{0}(x,y)&=&\sum_{i=-\infty}^{+\infty}{\phi^{(n)}_{0i}(x)
\phi^{(n)\dagger}_{0i}(y)\over E^{0}_{i}},\nonumber\\
\bar{\eta}^{'}&=&\bar{\eta}\exp[e\,\phi\,\gamma_{5}]h^{-1}\nonumber,\\
{\eta}^{'}&=&h\exp[e\,\phi\,\gamma_{5}]\eta,
\label{s63}
\eea

$E^{(0)}_{i}$ being the eigenvalues of $\hat{D}^{(n)}_{0}$ and
$\phi^{(n)}_{0i}(x)$ the related eigenfunctions: an explicit
expression for $S^{(n)}_{0}(x,y)$ is given in \cite{jay88}.

To calculate the generating functional for the fermionic fields we are
left with computing ${\det}^{'}[\hat{D}^{(n)}]$. The standard
$\zeta$--function calculation will give us a result that does not take
into account the Jackiw--Rajaraman ambiguity: in order to implement
correctly the freedom in the choice of the local term, carefully
considering the global meaning of the determinant, we use the usual
definition  \cite{Gro94}, generalized to the case of a
non--trivial connection:
\eq
\hat{\det}^{'}[\hat{D}^{(n)}]={{\det}^{'}[\hat{D}^{(n)}\hat{D}^{(n)}_{\alpha}]
\over {\det}^{'}[\hat{D}^{(n)}_{\alpha}]},
\label{s64}
\en

where
\eq
\hat{D}^{(n)}_{\alpha}=R\,\gamma_{a}e^{\mu}_{a}[\,iD_{\mu}+e\N_{\mu}+e\alpha
\hat{a}_{\mu}\,],
\label{s65}
\en

$\alpha$ being a real number. The parameter
ambiguity only affects the topologically trivial part of the gauge
connection: only in this way the operator eq.({\ref{s65}) is well defined
on $S^{2}$, because the  associated winding number is still an integer.
Moreover it is easy to prove that:
\bea
Ker[\hat{D}^{(n)}\hat{D}^{(n)}_{\alpha}]&=&Ker[\hat{D}^{(n)}_{\alpha}]
=n,\nonumber\\
Ker[\hat{D}^{(n)}_{\alpha}\hat{D}^{(n)}]&=&Ker[\hat{D}^{(n)}]=n.
\label{s66}
\eea

In our approach $\N_{\mu}$ is a classical field; the ambiguity of
regularization can only affect the quantum fluctuations, that depend on
quantum loops (determinant calculation): we expect that, with our
definition, the terms depending on $\alpha$ do not involve $\N_{\mu}$.
More precisely they must be local polynomials in the quantum
fluctuations and their derivatives.

The computation of eq.(\ref{s64}) is rather involved from the technical
point of view: essentially we have applied to the present situation the
theorems derived in \cite{mus83} and \cite{tho91},
to obtain (details are given in Appendix B)
\eq
{\det}^{'}[\hat{D}^{(n)}\hat{D}^{(n)}_{\alpha}]=
{\det}^{'}[(\hat{D}^{(n)}_{0})^{2}]\exp\int^{1}_{0} dt\,\omega^{'}(t)
\label{s67}
\en

and
\bea
\omega^{'}(t)&=&\inta
Tr\Bigl[\,K_{0}\Bigl(\,\hat{D}^{(n)}\hat{D}^{(n)}_{\alpha}(t);x,x\,\Bigr)
[e(1+\alpha)\phi\gamma_{5}+(1-\alpha)h(t)\partial_{t}h^{-1}(t)]+\nonumber\\
&+& K_{0}\Bigl(\,\hat{D}^{(n)}_{\alpha}\hat{D}^{(n)}(t);x,x\,\Bigr)
[e(1+\alpha)\phi\gamma_{5}-(1-\alpha)h^{-1}(t)\partial_{t}h(t)]\,
\Bigr]-\nonumber\\
&-&\sum_{m=1}^{|n|}\inta \varphi^{(n)\dagger}_{0m}(x,t)
[e(1+\alpha)\phi\gamma_{5}+(1-\alpha)h(t)\partial_{t}h^{-1}(t)]\varphi^{(n)}
_{0m}(x,t)-\nonumber\\
&-&\sum_{m=1}^{|n|}\inta \chi^{(n)\dagger}_{0m}(x,t)
[e(1+\alpha)\phi\gamma_{5}-(1-\alpha)h^{-1}(t)\partial_{t}h(t)]\chi^{(n)}
_{0m}(x,t),
\label{s68}
\eea

where $K_{0}({\cal A};x,y)$ is the analytic continuation in $s=0$ of the
kernel of the $s$--complex power of the operator ${\cal A}$ \cite{See67},
\big(see eq.(\ref{g2})\big);
$h(t)$ interpolates along the $U(1)$--valued functions between $h$ and
the identity, (remember that $\pi_{2}(S^{1})=0$), and
\bea
\hat{D}^{(n)}(t)&=&\hat{D}^{(n)}(\hat{a}_{\mu}(t)),\nonumber\\
\hat{D}^{(n)}_{\alpha}(t)&=&\hat{D}^{(n)}_{\alpha}(\hat{a}_{\mu}(t)),
\label{s69}
\eea

\eq
\hat{a}_{\mu}(t)=t\,\epsilon_{\mu\nu}g^{\nu\rho}\partial_{\rho}\phi+
{1\over ie}h(t)\partial_{\mu}h^{-1}(t),
\label{s70}
\en

\bea
\hat{D}^{(n)}(t)\chi^{(n)}_{0m}(x,t)&=& 0,\nonumber\\
\hat{D}^{(n)}_{\alpha}(t)\varphi^{(n)}_{0m}(x,t)&=& 0,
\label{s71}
\eea

$\chi^{(n)}_{0m}(x,t)$, $\varphi^{(n)}_{0m}(x,t)$ being the orthonormal
bases of the kernels of the operators in eq.(\ref{s69}), smoothly
interpolating between $\hat{D}^{(n)}$ and $\hat{D}_{0}^{(n)}$ and between
$\hat{D}^{(n)}_{\alpha}$ and $\hat{D}^{(n)}_{0}$ respectively.

In the same way:
\eq
{\det}^{'}[\hat{D}^{(n)}_{\alpha}]={\det}^{'}[\hat{D}^{(n)}_{0}]
\exp\int^{1}_{0} dt\,\omega^{''}(t)
\label{s72}
\en

and
\bea
\omega^{''}(t)&=&\inta
Tr\Bigl[\,K_{0}\Bigl(\,\hat{D}^{(n)}_{\alpha}(t);x,x\,\Bigr)
[e\,\alpha\,\phi\,\gamma_{5}]\Bigr]\nonumber\\
&-&\sum_{m=1}^{|n|}\inta \varphi^{(n)\dagger}_{0m}(x,t)
[e\,\alpha\,\phi\,\gamma_{5}]\varphi^{(n)}
_{0m}(x,t).
\label{s73}
\eea

The trace of the heat--kernel coefficients can be easily performed,
as in the previous section; on the other hand
the computation of the integrals over the ``interpolating'' zero modes is more
involved and subtler; for this reason we give the final result, deferring the
technical procedure to the Appendix B:
\bea
\hat{\det}^{'}[D^{(n)}]&=&\exp\Bigl[\,{e^2\over
2\pi}\inta\phi\Delta\phi\,+\,{e^2\over
2\pi}\gamma\inta[-\phi\Delta\phi+{1\over
e^2}\partial_{\mu}h\partial^{\mu}h^{-1}]\,\Bigr]\nonumber\\
& &|\det B|^{-2}{\det}^{'} [\hat{D}^{(n)}_{0}].
\label{s74}
\eea

The determinant of the zero mode matrix disappears from the generating
functional. The parameter $\gamma$ is linked to $\alpha$ by:
\eq
\gamma={1\over 2}(1-\alpha)^{2}.
\label{s75a}
\en

We remark that eq.(\ref{s74}) exhibits an ambiguity only in the quantum
fluctuations and, with respect to them, is local:
\eq
\inta[-\phi\Delta\phi+{1\over
e^2}\partial_{\mu}h\partial^{\mu}h^{-1}]=\inta
\hat{a}_{\mu}\hat{a}^{\mu}.
\label{s75}
\en

Moreover it is quite natural to have in any sector the same ambiguity:
it is related to the quantum fluctuation and not to the classical
background.
The last point is the calculation of ${\det}^{'} [\hat{D}^{(n)}_{0}]$,
that we present in Appendix C. The result is:
\eq
{\det}^{'} [\hat{D}^{(n)}_{0}]=\exp[\,-4\zeta^{'}_{R}(-1)+{n^{2}\over
2}+|n|\log |n|!-\sum^{|n|}_{m=1}2m\log m].
\label{s76}
\en

\vskip 0.5truecm
\section{The fermionic propagator}

The generating functional (\ref{s54}) for $J_{\mu}=0$ becomes:
\bea
{\cal Z}[\eta,\bar{\eta}]&=&{\cal Z}_0^{-1}(\lb)
\sum_{n=-\infty}^{+\infty}\exp[-{\pi
n^2\over 2e^2 R^2}]\int {\cal D}\phi{\cal D}h \exp[\inta\int
d^{2}y\sqrt{g}\,R\,\bar{\eta}^{'}(x)S^{(n)}_{0}(x,y)\eta^{'}(y)]\nonumber\\
& &{\det}^{'}[\hat{D}^{(n)}_{0}]\exp\Bigl[{e^{2}\over 2\pi}\inta
[(1-\gamma)\phi\Delta\phi+{1\over
e^2}\partial_{\mu}h\partial^{\mu}h^{-1}]\,\Bigr]\nonumber\\
& &\exp\Bigl[-S_{Bos.}[\lb;\phi,h]\Bigr]
\Pi^{n}_{m=1}[\inta\sqrt{R}\,\chi^{(n)\dagger}_{0m}\eta^{'}]
[\inta\sqrt{R}\,\bar{\eta}^{'}\chi^{(n)}_{0m}].
\label{s77}
\eea

All the fermionic correlation functions of the theory can be derived: the
functional integration over $\phi$ and $h$ is gaussian, as in the flat
case, and the model is still constructed by means of free fields (on
curved background).
Chirality--violating correlation functions can be different from
zero only in the non--trivial winding number sectors, as in
the case of the vector Schwinger model: actually, in that model, this
feature survives the limit $R\rightarrow\infty$, changing in this way the
vacuum structure of the theory \cite{sch77}. In particular the vacuum fermionic
condensate is seen to be different from zero \cite{jay88}:
\eq
<\bar{\psi}\psi>={e\over 2\pi}{\exp[C]\over\sqrt{\pi}}
\label{s78}
\en

recovering, by a path--integral procedure, the well known operatorial
result \cite{Low71} ($C$ is the Euler--Mascheroni constant).
As a matter of fact the fermionic propagator on the sphere has a different
algebraic structure, correlations being now possible between $\bar{\psi}_{R}$
and $\psi_{R}$ and between $\bar{\psi}_{L}$ and $\psi_{L}$. The relevant
contributions come from the $n=\pm 1$ sectors, while the sector
$n=0$ leads to the
usual chirality conserving correlation.

We calculate the explicit
form of the propagator; in the limit for $R\rightarrow \infty$ we
recover the expression found in the theory on the plane.
As a particular case we get the vacuum
expectation value of the fermionic scalar density: at variance
with the eq.(\ref{s78}) we
find a vanishing result, confirming the conjecture of \cite{Gir86} about the
triviality of the vacuum in the chiral Schwinger model and showing how
the breaking of the gauge invariance completely changes the structure of
the theory.

In order to perform the calculation it is useful to write
\eq
{1\over ie}h\partial_{\mu}h^{-1}=\partial_{\mu}\beta,
\label{s79}
\en

what is possible, thanks to the triviality of $\pi_{2}(S_{1})$.
We change variable from $h$ to $\beta$ in eq.(\ref{s77}) and rescale
\bea
\phi &\rightarrow & ({1+\lb^2 \over 2})\phi,\nonumber\\
\beta &\rightarrow & ({1+\lb^2 \over 2})\beta,\nonumber\\
e^{2} &\rightarrow & (1+\lb^2)^{2}{e^2 \over 4\pi}=\hat{e}^{2}.
\label{s80}
\eea

We use the notation:
\eq
\psi=\left( \begin{array}{cc} & \psi_{R}\\
                           & \psi_{L} \end{array}
                                                 \right),
\en
\eq
\bar{\psi}=(\bar{\psi}_{R},\bar{\psi}_{L}).
\label{s81}
\en

We can easily derive from the generating functional the fermionic propagator:

\eq
<\psi(x)\bar{\psi}(y)>=
\left( \begin{array}{cc}  S_{RR}(x,y)  &  S_{RL}(x,y) \\
                                      S_{LR}(x,y) &  S_{LL}(x,y)  \end{array}
                                                              \right);
\en
where
\bea
S_{RL,LR}(x,y)&=&S^{0}_{RL,LR}(x,y)\exp[-{1\over 2}G_{\pm}(x,y)]\\
S_{RR,LL}(x,y)&=&{1\over 4\pi R}\exp[{1\over 2}-{\pi\over 2 e^2
R^2}]\exp[{1\over 2}G_{R,L}(x,y)]
\label{s82}
\eea
\eq
\exp[-{1\over 2}G_{\pm}(x,y)]=
{\int {\cal D}\phi{\cal D}\beta \exp[-{1\over
2}\Gamma(\phi,\beta)\pm\sqrt{\pi}\hat{e}(\phi(x)-\phi(y))+
i\sqrt{\pi}\hat{e}(\beta(x)-\beta(y))]
\over
\int {\cal D}\phi{\cal D}\beta \exp[-{1\over
2}\Gamma(\phi,\beta)]},
\label{s83}
\en
\eq
\exp[-{1\over 2}G_{R,L}(x,y)]=
{\int {\cal D}\phi{\cal D}\beta \exp[-{1\over
2}\Gamma(\phi,\beta)\mp\sqrt{\pi}\hat{e}(\phi(x)+\phi(y))+
i\sqrt{\pi}\hat{e}(\beta(x)-\beta(y))]
\over
\int {\cal D}\phi{\cal D}\beta \exp[-{1\over
2}\Gamma(\phi,\beta)]}.
\label{s84}
\en
\eq
\Gamma(\phi,\beta)=\inta[\phi\Delta^{2}\phi+
\lb^{2}\beta\Delta^{2}\beta-2\lb\phi\Delta^{2}\beta-\hat{e}^{2}(1-\gamma)
\phi\Delta\phi+\hat{e}^{2}\gamma\beta\Delta\beta].
\label{s84a}
\en
The explicit form of the zero mode of $n=\pm 1$ \cite{bass} was taken into
account; $S^{0}_{RL,LR}(x,y)$ are the nonvanishing components
of the Green function
defined in eq.(\ref{s63}) that takes a very simple form in stereographical
coordinates \cite{jay88}:
\eq
S^{0}(\hat{x},\hat{y})={1\over 4\pi R}(1+{\hat{x}^2\over 4 R^2})^{1\over 4}
(1+{\hat{y}^2\over 4 R^2})^{1\over 4}{\gamma_{a}(\hat{x}_{a}-\hat{y}_{a})\over
(\hat{x}-\hat{y})^2}.
\label{s85}
\en

Now the integration over $\phi$ and $\beta$ is quadratic and can be
easily performed expanding, for example, the scalar fields in
spherical harmonics, that are (up to a scale factor) a complete set of
orthogonal eigenfunctions for the laplacian. No
problem arises with the zero mode thanks to the properties:
\bea
-\Delta\phi&=&f_{01}\nonumber\\
\inta\epsilon^{\mu\nu}f_{\mu\nu} &=&0
\label{s86}
\eea
\bea
\Delta\beta&=&D_{\mu}(h\partial^{\mu}h^{-1})\nonumber\\
\inta D_{\mu}(h\partial^{\mu}h^{-1})&=&0.
\label{s87}
\eea
The final result is:
\bea
G_{\pm}(\omega)&=&G^{U.V.}_{\pm}+\hat{G}_{\pm}(\omega)\nonumber\\
G_{R,L}&=&G^{U.V.}_{R,L}+\hat{G}_{R,L}(\omega)
\label{s88}
\eea
\eq
G^{U.V.}_{\pm}=\lim_{\alpha\rightarrow 0}
{1\over 2}{(1\mp i\lb)^{2}\over [\gamma(1+\lb^{2})-\lb^{2}]}
\sum_{l=1}^{\infty}{(2l+1)\over [l(l+1)+m^{2}(\lb)R^2]}
P_{l}(\cos\alpha),
\label{s89}
\en
\bea
\hat{G}_{\pm}(\omega)&=&{m^{2}(\lb) R^2\over 2(1-\gamma)\gamma}
\sum_{l=1}^{\infty}{(2l+1)\over l(l+1)[l(l+1)+m^{2}(\lb)R^2]}
[1-P_{l}(\cos\theta))]
\nonumber\\
&-&{1\over 2}{(1\mp i\lb)^{2}\over [\gamma(1+\lb^{2})-\lb^{2}]}
\sum_{l=1}^{\infty}{(2l+1)\over [l(l+1)+m^{2}(\lb)R^2]}
P_{l}(\cos\theta),
\label{s90}
\eea
\eq
G^{U.V.}_{R,L}=\lim_{\alpha\rightarrow 0}
{1\over 2}{(1-\lb^{2})\over [\gamma(1+\lb^{2})-\lb^{2}]}
\sum_{l=1}^{\infty}{(2l+1)\over [l(l+1)+m^{2}(\lb)R^2]}
P_{l}(\cos\alpha),
\label{s91}
\en
\bea
\hat{G}_{R,L}(\omega)&=&{m^{2}(\lb) R^2\over 2\gamma}
\sum_{l=1}^{\infty}{(2l+1)\over l(l+1)[l(l+1)+m^{2}(\lb)R^2]}
[1-P_{l}(\cos\theta))]
\nonumber\\
&+&{m^{2}(\lb) R^2\over 2(1-\gamma)}
\sum_{l=1}^{\infty}{(2l+1)\over l(l+1)[l(l+1)+m^{2}(\lb)R^2]}
[1+P_{l}(\cos\theta))]\nonumber\\
&+&{1\over 2}{(1+\lb^{2})\over [\gamma(1+\lb^{2})-\lb^{2}]}
\sum_{l=1}^{\infty}{(2l+1)\over [l(l+1)+m^{2}(\lb)R^2]}
P_{l}(\cos\theta),
\label{s92}
\eea
$P_{l}$ being the Legendre polynomials and $\omega$ the angle between
$\hat{r}(x)$ and $\hat{r}(y)$, the three--vectors representing the
points $x$ and $y$ on $S^{2}$, embedded in $R^{3}$; the mass $m^{2}(\lb)$ is
\eq
m^{2}(\lb)={\hat{e}^{2}\gamma(\gamma-1) \over
\gamma(1+\lb^{2})-\lb^{2} }.
\label{s93}
\en
In order to perform the limit on the plane we require that:
\[ \lim_{R\rightarrow\infty}Z(\lb=ir)= Z,\]
the effective action obtained in eq.(\ref{26}); this entails a
relation between $\gamma$ and $a$:
\eq
\gamma=-{a\over (1+\lb^2)}.
\label{s94}
\en

We recognize that
\[m^{2}(\lb)=m^2,\]
$m^2$ being the bosonic mass we have found in the calculation on the
plane.

Actually the series defining $G_{\pm}$ and $G_{R,L}$ can be summed in terms of
special functions. We can choose the point $y$ as the north pole,
$\cos\omega=\cos\theta$
without losing any information. Then $\hat{G}(\omega)$ depends only on the
polar
angle $\theta$: following \cite{jay88} we introduce the function
\eq
\Delta_{1}(\theta)=\sum_{l=1}^{\infty}
{(2l+1)P_{l}(\cos\omega)\over l(l+1)+m^2 R^2}+{1\over m^2 R^2}.
\label{s95}
\en

that satisfies the differential equation
\eq
[-{1\over r^2 \sin\theta}{\partial\over\partial\theta}
(\sin\theta{\partial\over\partial\theta})+m^2 ]\Delta_{1}(\theta)=0
\label{s96}
\en

for $\theta\neq 0$. The solution are the associated Legendre
functions \cite{bat}, usually denoted by $Q_{\nu}(x)$, $x=\cos\theta$. It
is not difficult to prove that with the usual normalization for $Q_{\nu}(x)$:
\bea
\Delta_{1}(\theta)&=&Q_{\nu_{1}}(\cos\theta)+Q_{\nu_{2}}(\cos\theta),
\nonumber\\
\nu_{1,2}&=&-{1\over 2}\pm\sqrt{{1\over 4}-m^2 R^2},
\label{s97}
\eea

obtaining
\eq
\sum_{l=1}^{\infty}
{(2l+1)P_{l}(\cos\omega)\over l(l+1)+m^2 R^2}=
\Bigl[\Bigl(Q_{\nu_{1}}(\cos\theta)+Q_{\nu_{2}}(\cos\theta)\Bigr)-
{1\over  m^2 R^2}\Bigr].
\label{s98}
\en
Then we use the following result \cite{bat}:
\eq
\sum_{l=1}^{\infty}
{(2l+1)P_{l}(\cos\omega)\over l(l+1)}=-\ln({1-\cos\theta\over 2})-1,
\label{s99}
\en
to get
\bea
\sum_{l=1}^{\infty}{m^2 R^2
(2l+1)\over l(l+1)[l(l+1)+m^{2}(\lb)R^2]}P_{l}(\cos\theta)&=&
-\Bigl[Q_{\nu_{1}}(\cos\theta)+Q_{\nu_{2}}(\cos\theta)\nonumber\\
&+&
\ln({1-\cos\theta\over 2})-{1\over m^2R^2}+1\Bigr]
\label{s100}
\eea
and
\eq
\sum_{l=1}^{\infty}{m^2 R^2
(2l+1)\over l(l+1)[l(l+1)+m^{2}(\lb)R^2]}=
\Bigl[\psi(1+\nu_{1})+\psi(1+\nu_{2})+2C+{1\over m^2R^2}-1\Bigr].
\label{s101}
\en
Eqs.(\ref{s89}-\ref{s92}) become
\eq
G^{U.V.}_{\pm}=
-{1\over 2}{(1\pm r)^{2}\over [a-r^2]}
\lim_{\alpha\rightarrow 0}\Bigl[\Bigl(Q_{\nu_{1}}(\cos\alpha)+
Q_{\nu_{2}}(\cos\alpha)\Bigr)-{1\over  m^2 R^2}\Bigr]
\label{s102}
\en
\bea
\hat{G}_{\pm}(\omega)&=&{1\over 2}{(1\pm r)^2(a-r^2-r)^2\over
a(a+1-r^2)(a-r^2)}[Q_{\nu_{1}}(\cos\theta)+Q_{\nu_{2}}(\cos\theta)
\Bigr]
\nonumber\\
&+&{1\over 2}{(1+r^{2})^2\over a(a+1-r^2)}\ln({1-\cos\theta\over2})
\nonumber\\
&+&{1\over 2}{(1-r^{2})^2\over a(a+1-r^2)}
\Bigl[\psi(1+\nu_{1})+\psi(1+\nu_{2})+2C\Bigr],
\label{s103}
\eea
\eq
G^{U.V.}_{R,L}=
{1\over 2}{(1+r^{2})\over a-r^2}\lim_{\alpha\rightarrow 0}
\Bigl[\Bigl(Q_{\nu_{1}}(\cos\alpha)+
Q_{\nu_{2}}(\cos\alpha)\Bigr)-{1\over  m^2 R^2}\Bigr],
\label{s104}
\en
\bea
\hat{G}_{R,L}(\omega)&=&-{1\over 2}{(1-r^{2})\over a-r^2}
\Bigl[Q_{\nu_{1}}(\cos\theta)+
Q_{\nu_{2}}(\cos\theta)\Bigr]
\nonumber\\
&-&{1\over 2}{(1-r^{2})^2\over a(a+1-r^2)}
\Bigl[\psi(1+\nu_{1})+\psi(1+\nu_{2})+2C\Bigr]\nonumber\\
&+&{1\over 2}{(1-r^{2})(2a+1-r^2)\over a(a+1-r^2)}
\Bigl[Q_{\nu_{1}}(\cos\theta)+
Q_{\nu_{2}}(\cos\theta)+\ln({1-\cos\theta\over 2})\Bigr].
\label{s105}
\eea
The term $G^{U.V.}_{\pm}$ and $G^{U.V.}_{R,L}$ are obviously related to the
ultraviolet divergencies we have found in the flat calculation:
they characterize
the short--range behaviour of the theory and therefore are present no matter
the global topology. One can directly check the independence from $R$
of the divergent
part. Moreover the relation
\eq
G^{U.V.}_{R,L}={1\over 2}[G^{U.V.}_{+}+G^{U.V.}_{-}],
\label{s106}
\en
is consistent with the redefinition
\bea
\psi^{Ren.}_{R}&=&Z^{-{1\over 2}}_{R}\psi_{R}\nonumber\\
\psi^{Ren.}_{L}&=&Z^{-{1\over 2}}_{L}\psi_{L},
\label{s107}
\eea
\bea
Z^{-{1\over 2}}_{R}&=&\exp [-{1\over 2}G^{U.V.}_{+}],\nonumber\\
Z^{-{1\over 2}}_{L}&=&\exp [-{1\over 2}G^{U.V.}_{-}],
\label{s108}
\eea
that leads to a meaningful expression for the correlation functions:
\bea
S^{Ren.}_{RL,LR}(\theta)&=&S^{0}_{RL,LR}(\theta)\exp[-{1\over 2}\hat{G}_{\pm}],
\nonumber\\
S^{Ren.}_{RR,LL}(\theta)&=&{1\over 4\pi R}\exp[{1\over 2}-{\pi\over 2 e^2
R^2}]\exp[-{1\over 2}\hat{G}_{R,L}].
\label{s109}
\eea

We expect that the small distance behaviour of the theory is the same of the
one in the flat case: let us study the limit
for $\theta\rightarrow 0$ in eq.(\ref{s109}).
The fermionic Green function in this limit has the expression
\[S^{0}_{RL,LR}={1\over 4\pi R}{1\over \theta},\]
that exhibits the canonical scaling in $R\theta$; the small--$\theta$
expansions of the exponents lead to
\bea
<\psi^{Ren.}_{R,L}(0)\bar{\psi}^{Ren.}_{L,R}(\theta)>&=&(R\theta)^{-[1+{1\over
2}{(1\pm r)^2\over a-r^2}]},\nonumber\\
<\psi^{Ren.}_{R,L}(0)\bar{\psi}^{Ren.}_{R,L}(\theta)>&=&(R\theta)^{{1\over
2}{1-r^2\over a-r^2}}.
\label{s110}
\eea

We notice that the scaling exponent of the chirality--conserving part of the
fermionic correlation function is exactly the same as in the flat case: we have
indeed the correct singularity as $\theta\rightarrow 0$. The contribution of
the $n=\pm 1$ sectors agree with our intuitive arguments for $r^2<1$: the
chirality--violating part goes to zero as $\theta$, in this range. For $r^2>1$
a singularity arises, changing drastically the analytical structure of
ultraviolet limit. In the very special case of $r^2=1$ we have:
\[<\psi^{Ren.}_{R,L}(0)\bar{\psi}^{Ren.}_{R,L}(\theta)>=
{1\over 4\pi R}\exp[{1\over 2}-{\pi\over 2 e^2 R^2}].\]

By the way there is another potential singularity in
eqs.(\ref{s103},\ref{s105}): the Legendre
function $Q_{\nu}$ has branch points at $x=\pm 1$ and $x=\infty$, so for
$\theta=\pi$ one could expect some critical behaviour. A careful use of the
asymptotic expansion near $x=-1$ \cite{bat}:
\[Q_{\nu}(x)={1\over 2}\cos(\nu\pi)\ln({1+x\over 2})+ O(1)\]
shows that the singularities cancel and the correlation function goes to a
constant value in this limit.

At this point we try to decompactify the theory: we have essentially to discuss
the large $R$ behaviour of two functions:
\[Q_{\nu_{1}}(\cos\theta)+
Q_{\nu_{2}}(\cos\theta)\]
and
\[\psi(1+\nu_{1})+\psi(1+\nu_{2}).\]
In order to check the flat--space limit let us define the geodesic distance:
\eq
\rho=R\theta
\label{s120}
\en

and let $R\rightarrow \infty $, keeping $R\theta$ fixed.
We define $\tau=\sqrt{m^2 R^2-{1\over 4}}$ which is real for
large $R$. Now:
\eq
Q_{\nu_{1}}(\cos\theta)+
Q_{\nu_{2}}(\cos\theta)=
\Bigl[{\pi\over \cosh(\pi\tau)}
P_{-{1\over 2}+i\tau}(-\cos\theta)-{1\over m^2 R^2}\Bigr].
\label{s121}
\en

An asymptotic expansion for large $\tau$ and small $\theta$ can be
found \cite{bat}:
\eq
P_{-{1\over
2}+i\tau}(-\cos\theta)={1\over \pi}\exp[\tau\pi]K_{0}(\tau\theta)+{\rho\over
R}+O({\rho\over R})^{2}.
\label{s122}
\en

Furthermore:
\bea
\tau\theta&=&\sqrt{m^2}\rho-{\rho\over 8\sqrt{m^2}}{1\over
R^2}\nonumber\\
K_{0}(\tau\theta)&=&K_{0}(\sqrt{m^2}\rho)+K_{1}(\sqrt{m^2}\rho)
{\rho\over 8\sqrt{m^2}}{1\over
R^2}+O({1\over R^4}),
\label{s123}
\eea

leading, in the limit $R\rightarrow\infty$, to
\eq
Q_{\nu_{1}}(\cos\theta)+
Q_{\nu_{2}}(\cos\theta)=
=2K_{0}(\sqrt{m^2}\rho)+O({1\over R^2}).
\label{nn}
\en

In the same way we use the asymptotic expansion for $\psi(z)$
($|z|\rightarrow\infty$):
\eq
\psi(z)=\ln z-{1\over 2z}+O({1\over z^2}),
\en
leading to
\eq
\psi({1\over 2}+i\tau)+\psi({1\over 2}-i\tau)=2\ln(mR)+O({1\over R^2}).
\label{mm}
\en

The first point to check is that the renormalization constants coincide in the
limit with the ones in the flat case: in order to compare the two
expressions we perform
the limit $\rho\rightarrow 0$ in eqs.(\ref{s103},\ref{s105}), after the
decompactification. We get:
\eq
Z_{R,L}=\lim_{\rho\rightarrow 0} \exp\Bigl[-{(1\pm r)^2\over2( a-r^2)}
K_{0}(\sqrt{m^2}\rho)\Bigr],
\label{s124}
\en
that corresponds to eq.(\ref{226}).
The renormalized two--point function becomes:
\bea
<\psi^{Ren.}_{R,L}(0)\bar{\psi}^{Ren.}_{L,R}(\rho)>&=&S^{0}(\rho)
\exp\Bigl[-{1\over 4}{(1-r^2)^2\over
a(a+1-r^2)}\ln(\tilde{m}^2\rho^2)\nonumber\\
&+&{(1\pm r)^2(a-r^2\pm r)^2\over
2a(a+1-r^2)(a-r^2)}K_{0}(\sqrt{m^2}\rho)\Bigr] +O({1\over R^2}),
\label{s125}
\eea
that coincides with eq.(\ref{224}), and
\eq
<\psi^{Ren.}_{R,L}(0)\bar{\psi}^{Ren.}_{R,L}(\rho)>={1\over 4\pi}R^{-(1+g)}
\exp[{1\over 2}F(\rho)],
\label{s126}
\en
\bea
F(\rho)&=&
\exp\Bigl[ {-1\over 4}{(1-r^2)^2\over
a-r^2}K_{0}(\sqrt{m^2}\rho)-{1-r^2\over 2 a}\ln(\tilde{m}^2\rho^2)\nonumber\\
&+&{(1-r^2)(1+2a-r^2)\over
a(a+1-r^2)}[K_{0}(\sqrt{m^2}\rho)+{1\over 2}\ln(\tilde{m}^2\rho^2)].
\label{s127}
\eea
The $n=0$ sector reproduces in the large $R$ limit the result of the
flat case. The
behaviour of the chirality--violating part is very different: it is
tuned
by the effective Thirring coupling $g$, we have found to describe the infrared
regime of the fermionic operators. In the first unitarity region,
where no ghost
is present, it turns out to be $g>-1$,
as it should for the consistency of the Thirring theory \cite{Col75}.
The contribution of $n=\pm 1$ is therefore suppressed
as $R\rightarrow\infty$; the
situation is very different from the Schwinger model, where
topological sectors contribute.

We expect a similar result for the vacuum expectation value of the fermionic
scalar density: we go back to eqs.(\ref{s91},\ref{s92}) and take the
limit $\theta\rightarrow0$
\eq
<\bar{\psi}\psi> ={\exp[{1\over 2}-{\pi\over 2 e^2R^2}]\over 2\pi R}
\exp{1\over 2}\Bigl[H^{U.V.}+H\Bigr],
\label{s128}
\en
where
\bea
H^{U.V.}&=&\lim_{\theta\rightarrow 0}{r^2\over a-r^2}[Q_{\nu_1}(\cos \theta)+
Q_{\nu_2}(\cos \theta)-{1\over m^2 R^2}],\nonumber\\
H &=&{1-r^2\over a+1-r^2}[\psi(1+\nu_1)+
\psi(1+\nu_2)+2C-1+{1\over m^2 R^2}].
\label{s129}
\eea
$H^{U.V.}$ is not a finite quantity, it is the ultraviolet divergence
related
to the renormalization constant for the composite operator $\bar{\psi}\psi$,
found in the flat space. It is very easy to prove, using the large $R$
expansion, that
\[Z=\exp\Bigl[{1\over 2}H^{U.V.}\Bigr]\]
coincides with eq.(\ref{236}).
After the renormalization:
\eq
<(\bar{\psi}\psi)_{Ren.}>={\exp[{1\over 2}-{\pi\over 2 e^2R^2}]\over 2\pi R}
\exp\Bigl[{1\over 2}H\Bigr].
\label{s130}
\en

If we perform the limit $R \rightarrow\infty$, we end up with:
\bea
<(\bar{\psi}\psi)_{Ren.}>&=&\lim_{R\rightarrow\infty}
{ R^{\delta_{1}}
\over
2\pi R }m^{\delta_{1}}\exp[\delta_{1} C+\delta_{2}],\nonumber\\
\delta_{1}&=&{(1-r^2)\over (a+1-r^2)},\nonumber\\
\delta_{2}&=&{a\over (a+1-r^2)}
\label{105}
\eea

The power of $R$ is:
\eq
R^{-\delta_{2}};
\label{106}
\en

for $a>r^2$ (in the first unitarity region, where no ghost is present) the
limit $R\rightarrow\infty$ is zero. The vacuum expectation value of the
scalar density vanishes for all the generalized chiral Schwinger models, in
the first unitarity region: in particular for the chiral Schwinger
model, confirming the conjecture proposed in \cite{Gir86}.

We end this section with some remarks concerning the bosonic spectrum:
as we have previously pointed out, the bosonic Green functions take
contribution only from the $n=0$ sector. The calculation of the propagator
for $A_{\mu}$ is straightforward and it can be expressed in terms of the
functions $Q_{\nu}(\cos\theta)$, taking also eq.(\ref{s99}) into
account; in the limit of
large $R$ we recover exactly the propagator we have found in the flat
case.

\vskip 0.5truecm

\section{Conclusions}

In conclusion we have thoroughly studied on the two-sphere $S^2$
a vector--axial vector theory
characterized by a parameter which interpolates between pure
vector and chiral Schwinger models: the generalized chiral Schwinger model.
The theory has been completely solved by means of non perturbative
techniques, obtaining explicit expressions for its correlators.

We have defined the theory, respecting its global character, a non--trivial
task because it was an anomalous one on a non--trivial
principal bundle. At least at our knowledge, it is the first time that
an anomalous gauge theory is quantized on a non--trivial topology,
taking into account the contributions of winding numbers different
from zero and zero--modes of the relevant fermionic operator.

We have discussed the definition of the Dirac--Weyl
determinant on $S^{2}$, in presence of topological charged gauge
connections, showing the appearance, in an analytical approach, of the
well--known fixed background connection of the cohomological solution
for the anomaly. We have explained its physical meaning and we have
carefully computed the Green's functions generating functional.

The bosonic spectrum is the same while parity--violating fermionic
correlators are seen to be different from zero for a  finite radius
of the sphere; ultraviolet
divergencies are present but their character is the same as in the flat
limit.

As an
application we have calculated the fermionic propagator and the
fermionic vacuum condensate: in the
flat--limit, the latter vanishes, at variance with the
behaviour in the vector Schwinger
model: no vacuum degeneracy is present in our general case, confirming the
very different structure between an anomalous (but still unitary) theory
and a gauge invariant one.
\vskip 0.5truecm
\underbar{ACKNOWLEDGEMENTS}

This work is carried out in the framework of the European Community
Research Programme ``Gauge theories, applied supersymmetry and quantum
gravity" with a financial contribution under contract SC1-CT92-D789.

\vskip 0.5truecm

\section{Appendix A}
In this appendix we discuss the dependence of our results on the zero
mode matrix eq.(\ref{s61bis}): in the standard gauge invariant situation,
the regularized product of the non vanishing eigenvalues develops a term
depending on the zero modes themselves \cite{ho79}. This term induces
a non local and non linear self--interaction on the gauge
fields, seemingly jeopardizing the solubility of the theory.
In ref.\cite{tho91} the decoupling of this complicated interaction was
proved on
the correlation function: it turns out that it can be expressed as
the determinant of the inverse matrix of the zero modes, appearing as an
effect of the source term in the generating functional eq.(\ref{s62}).
Our problem is slightly different: in order to introduce the
Jackiw--Rajaraman parameter, we use a gauge--non invariant definition of
the fermionic determinant eq.(\ref{s64}), so that an explicit calculation is
needed to prove the decoupling of such a kind of interaction from the
correlation functions.

We have essentially to compute the ratio:
\eq
{\exp[{\cal B}_{1}]\over \exp[{\cal B}_{2}]},
\label{b1}
\en

where
\bea
{\cal B}_{1}&=&-\sum_{m=1}^{|n|}\int_{0}^{1} dt
\inta \varphi^{(n)\dagger}_{0m}(x,t)
[e(1+\alpha)\phi\gamma_{5}+(1-\alpha)h(t)\partial_{t}h^{-1}(t)]\varphi^{(n)}
_{0m}(x,t)-\nonumber\\
&-&\sum_{m=1}^{|n|}\int_{0}^{1} dt \inta \chi^{(n)\dagger}_{0m}(x,t)
[e(1+\alpha)\phi\gamma_{5}-(1-\alpha)h^{-1}(t)\partial_{t}h(t)]\chi^{(n)}
_{0m}(x,t),
\label{b2}
\eea

and
\bea
{\cal B}_{2}&=&-\sum_{m=1}^{|n|}\int_{0}^{1} dt
\inta \chi^{(n)\dagger}_{0m}(x,t)
[e\,\alpha\,\phi\,\gamma_{5}]\chi^{(n)}
_{0m}(x,t),
\label{b3}
\eea

(see eq.(\ref{s68}) and eq.(\ref{s73}) respectively). It turns out to be
that:
\bea
{\cal B}_{1}-{\cal B}_{2}&=&{\cal D}_{1}+{\cal D}_{2}\nonumber\\
{\cal D}_{1}&=&-\sum_{m=1}^{|n|}\int_{0}^{1} dt
\inta \varphi^{(n)\dagger}_{0m}(x,t)
[2e\phi\,\gamma_{5}]\varphi^{(n)}
_{0m}(x,t),\nonumber\\
{\cal D}_{2}&=&
-\sum_{m=1}^{|n|}\int_{0}^{1} dt
\inta \varphi^{(n)\dagger}_{0m}(x,t)
[e(1-\alpha)(\phi\gamma_{5}+h(t)\partial_{t}h^{-1}(t))]\varphi^{(n)}
_{0m}(x,t)+\nonumber\\
&+&\sum_{m=1}^{|n|}\int_{0}^{1} dt \inta \chi^{(n)\dagger}_{0m}(x,t)
[e(1-\alpha)(\phi\gamma_{5}+h^{-1}(t)\partial_{t}h(t))]\chi^{(n)}
_{0m}(x,t).
\label{b4}
\eea

Let us consider firstly ${\cal D}_{1}$: from the orthonormality of
$\chi^{(n)}_{0m}(x,t)$ we get
\eq
\sum_{m=1}^{n}\sum_{k=1}^{n}B_{im}(t)B_{jk}(t)\inta\chi^{(n)\dagger}_{0k}(x)
\exp[-2e\,t\phi\,\gamma_{5}]\chi^{n}_{0m}(x)=\delta_{ij},
\label{b5}
\en

where we have defined the interpolating matrix:
\eq
\chi^{(n)}_{m}(x,t)=u\exp[-e\,t\phi\gamma_{5}]\sum_{j=1}^{n}B_{mj}(t)
\chi^{(n)}_{0j}(x),
\label{b6}
\en
\bea
B(0)&=&{1\kern-.8mm \mbox{\rm l}},\nonumber\\
B(1)&=&B.
\label{b7}
\eea

Eq.(\ref{b6}) implies the relations:
\bea
B(t)C^{T}(t)B(t)&=&{1\kern-.8mm \mbox{\rm l}},\nonumber\\
|\det B(t)|^{2}&=&(\det C(t))^{-1},
\label{b8}
\eea

the matrix $C(t)$ being:
\eq
C_{ij}(t)=\inta
\chi^{(n)\dagger}_{0i}(x)\exp[-2e\,t\phi\,\gamma_{5}]\chi^{(n)}_{0j}(x).
\label{b9}
\en

We are able to express ${\cal D}_{1}$ in a very compact way:
\bea
{\cal D}_{1}&=&\sum_{m,k,j=1}^{n}\int_{0}^{1}dt B^{*}_{mj}(t)B_{mk}(t)
\inta\chi^{(n)\dagger}_{0j}(x)
\exp[-2e\,t\phi\,\gamma_{5}]\chi^{n}_{0k}(x),\nonumber\\
&=&Tr\Bigl[\int_{0}^{1}dt\,B(t){d\over
dt}\Bigl(C^{T}(t)\Bigr)B^{\dagger}(t)\Bigr].
\label{b10}
\eea

Eq.(\ref{b8}) leads to:
\bea
{\cal D}_{1}&=&Tr\Bigl[\int_{0}^{1}dt\,{d\over
dt}\Bigl(C^{T}(t)\Bigr)C^{-1}(t)\Bigr]\nonumber\\
&=&Tr\log C^{T}(1),
\label{b11}
\eea

giving
\eq
\exp {\cal D}_{1}=\det C(1)=|\det B|^{-2}.
\label{b12}
\en

This is the term appearing in eq.(\ref{s75}), that cancels the
contribution of the sources. We have now to prove the vanishing of
${\cal D}_{2}$; we define the matrix $E$:
\eq
\varphi^{(n)}_{0m}(x,t)=\sum^{n}_{m=1}E_{mk}(t)\exp[(1-\alpha)t
(e\phi\gamma_{5}-i\beta)]\,\chi^{(n)}_{0k}(x,t),
\label{b13}
\en

$\beta$ being related to the gauge dependent part of $\hat{a}_{\mu}$ in
eq.(\ref{s86}). We can easily verify the consistency of eq.(\ref{s14})
being
\eq
\exp[-(1-\alpha)t\,(e\phi-i\beta)]\varphi^{(n)}_{0m}(x,t)\in
Ker[\hat{D}^{(n)}(t)].
\label{b14}
\en

{}From the relations:
\eq
\inta \chi^{(n)\dagger}_{0j}(x,t)\exp[-(1-\alpha)t\,
(e\phi\,\gamma_{5}-i\beta)] \varphi^{(n)}_{0m}(x,t)=E_{mj}
\label{b15}
\en

and
\eq
\inta \varphi^{(n)\dagger}_{0j}(x,t)\exp[(1-\alpha)t\,
(e\phi\,\gamma_{5}+i\beta)] \chi^{(n)}_{0m}(x,t)=E^{*}_{mj},
\label{b15bis}
\en

we obtain:
\eq
\inta \varphi^{(n)\dagger}_{0j}(x,t)\exp[(1-\alpha)t\,
(e\phi\,\gamma_{5}+i\beta)] \varphi^{(n)}_{0m}(x,t)=Tr[E {d\over
dt}E^{-1}]^{*}
\label{b16}
\en

and
\eq
-\inta \chi^{(n)\dagger}_{0j}(x,t)\exp[(1-\alpha)t\,
(e\phi\,\gamma_{5}-i\beta)] \chi^{(n)}_{0m}(x,t)=D_{mj}=Tr[E^{-1}{d\over
dt}E ]^{*}.
\label{b17}
\en

Taking into account eq.(\ref{b16}) and eq.(\ref{b17}) we find the
desired
result:

\eq
{\cal D}_{2}=\int_{0}^{1}dtTr[{d\over dt}(EE^{-1})]^{*}=0.
\label{b18}
\en
\section{Appendix B}
In this appendix we sketch the procedure to get
\[\hat{\det}^{'}[\hat{D}^{(n)}]\]

as defined in eq.(\ref{s64}): we do not give the details of the computation,
relying on a careful use of the techniques derived in \cite{mus83} and
\cite{tho91}, and only show the relevant steps to arrive to
eqs.(\ref{s67},\ref{s68}) and eqs.(\ref{s72},\ref{s73}).

Let us define:
\eq
\hat{D}^{(n)}(t)=h(t)\exp[e\,t\phi\,\gamma_{5}]\hat{D}^{(n)}_{0}
\exp[e\,t\phi\,
\gamma_{5}]h^{-1}(t)
\label{appy1}
\en

and
\eq
\hat{D}^{(n)}_{\alpha}(t)=h(\alpha t)\exp[e\,\phi\,\alpha t\gamma_{5}]
\hat{D}^{(n)}_{0}\exp[e\,\alpha t\phi\,\gamma_{5}]h^{-1}(\alpha t).
\label{appy2}
\en

$h(t)$ interpolates along the $U(1)$--valued functions between $h$ and
the identity (remember that $\pi_{2}(S^{1})=0$). Then
\eq
\omega^{'}(t)={d\over dt}\ln\Bigl[{\det}^{'}
[\hat{D}^{(n)}\hat{D}^{(n)}_{\alpha}]
\Bigr],
\label{appy3}
\en

so that eq.(\ref{s67}) follows. We notice that the problem is now
reduced to compute a derivative, that is the infinitesimal variation of
a determinant in the present case, and to an integration over a real
parameter. The subtle point relies in the fact that one has to consider
only the non-vanishing eigenvalues in the $\zeta$--function definition
of the determinant. The correct procedure was found in \cite{mus83} where
the authors define:
\eq
\det[{\cal O}]=\lim_{\epsilon\rightarrow 0}{1\over \lb^{N}}\det[{\cal
O}+\epsilon\unity],
\label{appy4}
\en

$N$ being the dimension of the kernel. The applications of the theorems
developed there to our operator leads to:
\bea
\omega^{'}(t)&=&Tr\Bigl[[\hat{D}^{(n)}(t)\hat{D}^{(n)}_{\alpha}(t)]^{-s-1}
A_{1}]_{s=0}
\nonumber\\
&-&\inta
Tr\Bigl[[e(1+\alpha)\phi\gamma_{5}+
(1-\alpha)h(t)\partial_{t}h^{-1}(t)]P_{1}(x,x)\Bigr]\nonumber\\
&-& \inta Tr\Bigl
[[e(1+\alpha)\phi\gamma_{5}-(1-\alpha)h^{-1}(t)\partial_{t}h(t)]P_{2}(x,x)\Bigr],
\label{appy5}
\eea

where
\bea
A_{1}&=&\Bigl[[e(1+\alpha)\phi\gamma_{5}+h(t(1-\alpha)){d\over dt}
h^{-1}(t(1-\alpha))]\,,\hat{D}^{(n)}(t)\hat{D}^{(n)}_{\alpha}(t)\Bigr]_{+}
\nonumber\\
&+&\hat{D}^{(n)}(t)[e(1+\alpha)\phi\gamma_{5}-h(t(1-\alpha)){d\over dt}
h^{-1}(t(1-\alpha))]\hat{D}^{(n)}_{\alpha}(t)
\label{appy6}
\eea

and $P_{1}(x,x)$, $P_{2}(x,x)$ are respectively the projectors on the
kernel of $\hat{D}^{(n)}\hat{D}^{(n)}_{\alpha}$ and $\hat{D}^{(n)}_{\alpha}
\hat{D}^{(n)}$: the computation of the functional and algebraic traces
leads to eq.(\ref{s68}). Along the same line the calculation for
${\det}^{'}[\hat{D}^{(n)}]$ follows.

\vskip 0.5truecm

\section{Appendix C}
We report the explicit $\zeta$--function calculation of the determinant
in eq.(\ref{s76}): we feel that our procedure is clearer than the
original one, presented in \cite{ho79}.

The eigenvalue equation for the operator $\hat{D}^{(n)}_{0}$ gives the
result \cite{bass} (we take $n$ positive for sake of simplicity):

Eigenvalues:$\quad\quad\quad\pm\sqrt{l(l+1)},\quad\quad$ $l>0$

Multiplicity:$\quad\quad\quad 2l+1.$

The relevant $\zeta$--function is (we compute essentially
$\det^{'}[\hat{D}^{(n)}_{0}]^{2}$):
\eq
\zeta(s)=\sum_{l=1}^{\infty}\,(2l+n)[l(l+n)]^{-s}.
\label{c1}
\en

By performing a binomial expansion we get:
\bea
\zeta(s)&=&\sum_{l=1}^{\infty}\,\sum_{k=0}^{\infty}\,
{(-1)^{k}\over
k!}\,{\Gamma(s+k)\over \Gamma(s)}n^{k}l^{-s-k}l^{-s}(2l+n)\nonumber\\
&=&2\zeta_{R}(1+2s)+n(1-2s)\zeta_{R}(2s)+n^2 s^2
\zeta_{R}(1+2s)\nonumber\\
&+&s\sum_{k=3}^{\infty}\,{(-1)^{k}\over
k!}\,{\Gamma(s+k)\over \Gamma(s+1)} {2s+k-2\over
s+k-1}n^{k}\zeta_{R}(2s+k-1).
\label{c2}
\eea

The series converges for $s=0$: the first terms are defined by
analytic continuation ($\zeta_{R}(s)$):
\bea
F(s,n)&=&\sum_{k=3}^{\infty}\,{(-1)^{k}\over
k!}\,{\Gamma(s+k)\over \Gamma(s+1)} {2s+k-2\over
s+k-1}n^{k}\zeta_{R}(2s+k-1),\nonumber\\
\lim_{s\rightarrow 0}F(s,n)&=&F(0,n),
\label{c3}
\eea

and
\bea
\zeta^{'}(0)&=&4\zeta^{'}_{R}(-1)+[-2n\zeta_{R}(0)+2n\zeta^{'}_{R}(0)]+
\nonumber\\
&+& n^2{d\over ds}
\Bigl[s^2\Bigl({1\over 2s}-\psi(1)+O(s)\Bigr)\Bigr]_{s=0}+F(0,n),\nonumber\\
&=& 4\zeta^{'}_{R}(-1)+n-n\log 2\pi +{n^2\over 2}+F(0,n).
\label{c4}
\eea

Let us compute $F(0,n)$:
\bea
F(0,n)&=&\sum_{k=3}^{\infty}\,{(-1)^{k}\over
k!}\,{\Gamma(k)\over \Gamma(1)} {k-2\over
k-1}n^{k}\zeta_{R}(k-1)=\nonumber\\
&=&\sum_{k=3}^{\infty}\,{(-1)^{k}\over \Gamma[k-1]}{k-2\over k(k-1)}n^k
\int^{\infty}_{0} dt\,t^{k-2}{e^{-t}\over 1-e^{-t}}=\nonumber\\
&=&\sum_{k=3}^{\infty}\,
\int^{\infty}_{0} dt\,{e^{-t}\over 1-e^{-t}}\Bigl[t^{k-2}(k-1){(-1)^{k}\over
k!}
n^k  -t^{k-2}{(-1)^{k}\over k!}n^k\Bigr]=\nonumber\\
&=&\sum_{k=3}^{\infty}\,
\int^{\infty}_{0} dt\,{e^{-t}\over 1-e^{-t}}\Bigl[{d\over dt}\Bigl(
t^{k-1}{(-1)^{k}\over k!}
n^k\Bigr)  -t^{k-2}{(-1)^{k}\over k!}n^k\Bigr]=\nonumber\\
&=&
\int^{\infty}_{0} dt\,{e^{-t}\over 1-e^{-t}}\Bigl[
{d\over dt}\Bigl({e^{-nt}-1+nt-{1\over 2}n^2 t^2 \over t}\Bigr)-
{e^{-nt}-1+nt-{1\over 2}n^2 t^2 \over t^2}
\Bigr]=\nonumber\\
&=&
\int^{\infty}_{0} dt\,{e^{-t}\over 1-e^{-t}}\Bigl[
{2\over t^2}-2{e^{-nt}\over t^2}-{n\over t}-n{e^{-nt}\over t}\Bigr].
\label{c7}
\eea

It is very easy to verify the convergence of the integral for $t=0$.
Let us define the function:
\bea
G(x)&=&\int^{\infty}_{0} {dt\over t^2}\,{e^{-t}\over 1-e^{-t}}\Bigl[
2-2e^{-xt}-tx-xte^{-xt}\Bigr]\nonumber\\
{dG(x)\over dx}&=&\int^{\infty}_{0} {dt\over t}\,
{e^{-t}\over 1-e^{-t}}\Bigl[
-1+e^{-xt}+xte^{-xt}\Bigr]\nonumber\\
G(0)&=&0.
\label{c8}
\eea

All this implies that:
\eq
F(0,n)=\int_{0}^{n} dx {dG(x)\over dx}.
\label{c9}
\en

We compute the derivative:
\bea
{dG(x)\over dx}&=&\int_{0}^{\infty}{dt\over t}\Bigl[
{e^{-(1+x)t}-e^{-t} \over 1-e^{-t}}+xe^{-t}\Bigr]+\nonumber\\
&+&\int_{0}^{\infty}{dt\over t}\Bigl[
{xte^{-(1+x)t}\over 1-e^{-t}}-xe^{-t}\Bigr]=\nonumber\\
&=&\log\Gamma(1+x)+x\int_{0}^{\infty}{dt\over t}\Bigl[
{te^{-(1+x)t}\over 1-e^{-t}}-e^{-t}\Bigr]=\nonumber\\
&=&\log\Gamma(1+x)+x\int_{0}^{1}{dy\over \log y}\Bigl[
{\log y\,y^{x}\over 1-y}+1\Bigr]=\nonumber\\
&=&\log\Gamma(1+x)-x\psi(1+x).
\label{c10}
\eea

The remaining integration is not difficult \cite{bat}:
\bea
F(0,n)&=&\int_{0}^{n}dx\log\Gamma(1+x)-\int_{0}^{n}dx\,x\psi(1+x)=\nonumber\\
&=&\int_{0}^{n}dx\log\Gamma(1+x)-n\log\Gamma(1+n)\nonumber\\
F(0,n)&=&\sum_{k=1}^{n}2k\log k+n\log 2\pi-n-n^2-n\log n!.
\label{c11}
\eea

The final result is:
\eq
\zeta^{'}(0)=
4\zeta^{'}_{R}(-1)-{n^2\over 2}-n\log n+\sum_{k=1}^{n}2k\log k.
\label{c12}
\en

\vfill\eject

\vfill\eject

\end{document}